\documentclass[lettersize,journal]{IEEEtran}
\usepackage{amsmath,amsfonts}
\usepackage{algorithmic}
\usepackage{algorithm}
\usepackage{array}
\usepackage[caption=false,font=normalsize,labelfont=sf,textfont=sf]{subfig}
\usepackage{textcomp}
\usepackage{stfloats}
\usepackage{url}
\usepackage{verbatim}
\usepackage{graphicx}
\usepackage{caption}
\usepackage{booktabs}
\usepackage{cite}
\usepackage{hyperref}
\usepackage{color}
\begin{document}

\title{TrafficMoE: Heterogeneity-aware Mixture of Experts for Encrypted Traffic Classification}

\author{Qing He, Xiaowei Fu, and Lei Zhang,~\IEEEmembership{Senior Member,~IEEE}
        % <-this % stops a space
\thanks{This work was partially supported by National Natural Science Fund of China under Grants 92570110 and 62271090, Chongqing Natural Science Fund under Grant CSTB2024NSCQ-JQX0038, and National Youth Talent Project. \textit{(Corresponding author: Lei Zhang)}}% <-this % stops a space
\thanks{\IEEEcompsocthanksitem Q. He, X. Fu and L. Zhang are with the School of Microelectronics and Communication Engineering, Chongqing University, Chongqing 400044, China.
(E-mail: qinghe@cqu.edu.cn, leizhang@cqu.edu.cn, xwfu@cqu.edu.cn)}
\thanks{Manuscript received April 19, 2021; revised August 16, 2021.}}

\markboth{Journal of \LaTeX\ Class Files,~Vol.~14, No.~8, August~2021}%
{Shell \MakeLowercase{\textit{et al.}}: A Sample Article Using IEEEtran.cls for IEEE Journals}


\maketitle

\begin{abstract}
Encrypted traffic classification is a critical task for network security. While deep learning has advanced this field, the occlusion of payload semantics by encryption severely challenges standard modeling approaches. Most existing frameworks rely on static and homogeneous pipelines that apply uniform parameter sharing and static fusion strategies across all inputs. This ``one-size-fits-all" static design is inherently flawed: by forcing structured headers and randomized payloads into a unified processing pipeline,  it inevitably entangles the raw protocol signals with stochastic encryption noise, thereby degrading the fine-grained discriminative features. In this paper, we propose TrafficMoE, a framework that breaks through the bottleneck of static modeling by establishing a Disentangle–Filter–Aggregate (DFA) paradigm. Specifically, to resolve the structural between-components conflict, the architecture disentangles headers and payloads using dual-branch sparse Mixture-of-Experts (MoE), enabling modality-specific modeling. To mitigate the impact of stochastic noise, an uncertainty-aware filtering mechanism is introduced to quantify reliability and selectively suppress high-variance representations. Finally, to overcome the limitations of static fusion, a routing-guided strategy aggregates cross-modality features dynamically, that adaptively weighs contributions based on traffic context. With this DFA paradigm, TrafficMoE maximizes representational efficiency by focusing solely on the most discriminative traffic features. Extensive experiments on six datasets demonstrate TrafficMoE consistently outperforms state-of-the-art methods, validating the necessity of heterogeneity-aware modeling in encrypted traffic analysis. The source code is publicly available at \url{https://github.com/Posuly/TrafficMoE_main}.
\end{abstract}


\section{Introduction}
\IEEEPARstart{N}{etwork} traffic classification plays a fundamental role in network security, traffic management, and quality-of-service (QoS) assurance. With the rapid growth of the Internet and IoT deployments, modern network traffic has become increasingly complex and diverse. In particular, the widespread adoption of encryption technologies (e.g., TLS) and anonymization mechanisms (e.g., VPNs) has significantly challenged traditional traffic analysis techniques, making encrypted traffic classification a critical yet challenging problem.

Early studies mainly relied on port-based and payload-based methods for traffic classification \cite{P2P,DPI}. While these approaches can achieve satisfactory performance in plaintext scenarios, their effectiveness is severely degraded under encrypted traffic. To address this issue, statistical and machine learning–based methods were proposed, leveraging flow-level features such as packet sizes and temporal patterns \cite{RSA,website}. Although applicable to encrypted environments, these methods depend heavily on hand-crafted features and often suffer from limited generalization. More recently, deep learning based approaches have demonstrated superior capability in automatically learning discriminative representations from raw traffic data, yet they typically require large-scale labeled datasets and still struggle to generalize across unseen applications and encryption settings \cite{CNN1,RNN1,robustTIFS,ATVITSC,CBSeq}.
Inspired by the success of self-supervised learning, pre-training techniques have emerged as a promising paradigm for encrypted traffic classification \cite{Et-bert,trafficformer,EAPT}. By learning generic traffic representations from large amounts of unlabeled data, these approaches significantly reduce the reliance on labeled samples and improve robustness across domains. Despite these advances, most existing frameworks rely on largely homogeneous modeling pipelines. By employing uniform parameter sharing and static fusion strategies across the whole traffic data, these methods implicitly operate under a stationary assumption: \textit{treating the contribution of diverse traffic components across samples equally}. This one-size-fits-all design overlooks the complexity of encrypted traffic by simply forcing structurally diverse signals into a static processing pipeline. Consequently, the model lacks flexibility to varying information density inherent in different traffic segments.

However, \textit{encrypted traffic is inherently heterogeneous}. A typical flow comprises two distinct modalities: \textbf{headers}, which encapsulate deterministic protocol logic, and \textbf{encrypted payloads}, which exhibit high-entropy and stochastic characteristics. Moreover, the discriminative properties of these components are not uniform, which fluctuates depending on the application type and encryption settings. Modeling such dynamic data with static architectures creates a misalignment in inductive bias. Specifically, \textit{uniform mechanisms inherently conflate protocol logic with random encryption noise, thereby diminishing the model's sensitivity to fine-grained features}.

From this perspective, current limitations mainly arise from three key factors. %insufficient conditional specialization. 
\textit{First}, uniform parameter sharing constrains modeling capacity, failing to decouple the deterministic syntax of headers from the stochastic patterns of payloads. \textit{Second}, indiscriminate processing treats all tokens equally, allowing inherent encryption noise to propagate and degrade feature quality. \textit{Third}, static fusion strategies overlook the sample-specific context and the varying discriminative utility of headers as well as payloads across flows. Collectively, these issues facilitate the proposed new framework that explicitly disentangles semantic modeling of headers and encrypted payloads, and dynamically adapts feature integration.

To address this challenge, we propose \textbf{TrafficMoE}, a heterogeneity-aware framework designed to harmonize modeling architectures according to the heterogeneous properties of headers and encrypted payloads in the encrypted traffic. The framework follows a Disentangle–Filter–Aggregate paradigm. Specifically, headers and payloads are decoupled into separate branches, each employing sparse Mixture-of-Experts (MoE) to enable disciminative modeling. To counteract noisy components, an uncertainty-aware filtering mechanism is proposed to quantify token reliability and selectively attenuate high-variance patterns. Finally, a routing-guided fusion strategy dynamically re-weights cross-modality features according to sample-specific context. Through this paradigm, TrafficMoE shifts from the ``one-size-fits-all" static modeling to heterogeneity-aware dynamic modeling, strictly optimizing representation capacity for different traffic segments.

\begin{figure}[!t]
    \centering
    \vspace*{-2.5mm}
    \subfloat[Existing  paradigm\label{fig:em}]{
        \includegraphics[width=0.4\linewidth]{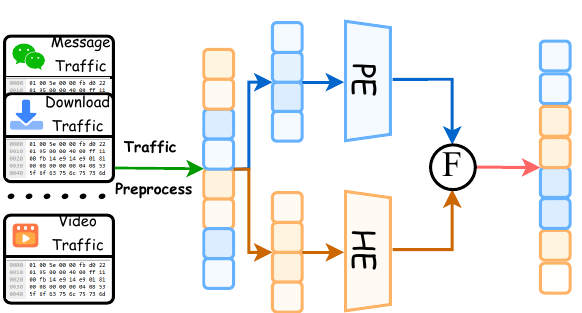}
    }\hfill
    \subfloat[ TrafficMoE framework\label{fig:om}]{
        \includegraphics[width=0.5\linewidth]{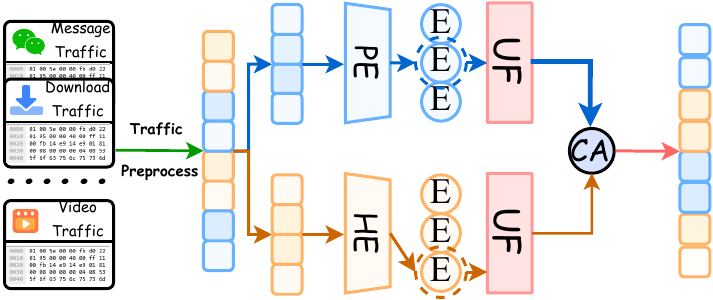}
    }
    \caption{Comparison between existing modeling paradigms and the proposed TrafficMoE framework. 
    Existing paradigm (a) typically process heterogeneous traffic components in a unified manner with static fusion strategies, whereas TrafficMoE (b) explicitly disentangles headers and payloads, incorporates uncertainty-aware filtering (UF), and performs conditional aggregation (CA) guided by MoE routing probabilities for adaptive and context-aware integration.}
    \label{fig:figure1}
\end{figure}

Fig.~\ref{fig:figure1} illustrates the key differences between existing traffic classification pipelines and our proposed framework. 
Fig.~\ref{fig:em} indicates that in existing paradigm, headers and payloads are typically processed as largely homogeneous sequences with static fusion strategies, that treat all traffic components equally. 
This design implicitly assumes uniform contribution from different components, potentially obscuring their individual discriminative representations. 
In contrast, Fig.~\ref{fig:om} indicate TrafficMoE explicitly disentangles the headers and payloads into separate modalities, applies an uncertainty-aware filtering (UF) to suppress unreliable tokens, and performs a routing guided implicit conditional aggregation (CA). %with MoE routing probabilities. 
This dynamic and context-aware integration allows the model to adaptively leverage heterogeneous traffic cues according to sample-specific characteristics, resulting in more effective and robust representation learning for headers and payloads.

% In summary, we make the following major contributions:
% \begin{itemize}
%     \item We propose TrafficMoE, a dual-branch Mixture-of-Experts framework that explicitly disentangles headers and payloads, enabling specialized modeling of heterogeneous traffic components and addressing the limitations of homogeneous modeling in existing approaches.
%     \item We introduce an uncertainty-aware filtering mechanism to suppress high-uncertainty tokens, thereby improving the reliability and robustness of learned representations under encrypted and diverse network traffic.
%     \item We design a conditional fusion strategy guided by MoE routing probabilities, allowing dynamic integration of heterogeneous features according to traffic context and sample-specific characteristics.
%     \item We conduct extensive experiments on multiple encrypted traffic datasets, demonstrating that TrafficMoE consistently outperforms state-of-the-art methods in terms of accuracy, generalization, and robustness across different datasets and encryption settings.

% \end{itemize}

In summary, our contributions are threefold:

\begin{itemize}

    \item We identify the uniform modeling of headers and payloads without distinction as a fundamental limitation of existing homogeneous pipelines, and reformulate a new encrypted traffic classification framework from a novel perspective of heterogeneity-aware disentangling modeling across heterogeneous traffic components and sample-dependent contexts. %This perspective emphasizes the necessity of disentangling modeling across heterogeneous traffic components and sample-dependent contexts.

    \item We propose TrafficMoE, a heterogeneity-aware sparse Mixture-of-Experts architecture that realizes hierarchical computation for headers and encrypted payloads. The framework follows a Disentangle–Filter–Aggregate paradigm, i.e., explicitly disentangles the headers and payloads as two modalities, filters unreliable tokens via uncertainty weighting and implicitly aggregate features across modalities for encrypted traffic representation. %aggregates modality-specific expert specialization, uncertainty-aware token filtering for reliability modeling, and routing-guided conditional fusion for adaptive cross-modality integration within a coherent disentangle–filter–conditionally aggregate paradigm.

    \item We conduct extensive experiments on six encrypted traffic datasets and demonstrate that TrafficMoE consistently achieves state-of-the-art classification performance compared with existing methods.

\end{itemize}

\section{Related Work}
\subsection{Encrypted Traffic Classification}

Encrypted traffic classification has been extensively studied in network security and traffic analysis, due to the widespread adoption of encryption protocols that invalidate traditional port-based and payload inspection techniques. Early learning-based approaches focused on extracting statistical features or modeling packet sequences to characterize encrypted flows. Deep Packet \cite{CNN1} leverages stacked autoencoders and convolutional networks to classify encrypted applications directly from raw packet bytes, demonstrating the feasibility of deep representation learning in encrypted scenarios.
Subsequent studies explored more expressive sequence modeling architectures to capture temporal and structural patterns within traffic flows. Flow-level sequence models, such as FS-Net \cite{FSNet2019}, treat encrypted traffic as ordered packet sequences and employ convolutional or recurrent architectures to learn discriminative flow representations. Hybrid CNN--RNN models further integrate spatial and temporal feature extraction, as exemplified by TSCRNN \cite{TSCRNN2021}, which is particularly effective in industrial IoT environments. In parallel, several works reformulate traffic as image-like types in order to exploit the maturity of 2D convolutional networks for encrypted traffic classification \cite{turn0search7}.

More recently, pretraining techniques have emerged as a dominant paradigm for universal representation. %generalization under limited labeled data. 
ET-BERT \cite{Et-bert} introduces transformer-based pretraining on large-scale unlabeled traffic corpora, learning contextualized universal representations transferable to multiple encrypted traffic tasks. Building upon this idea, generative or language-inspired models such as NetGPT \cite{turn0search2} and Language of Network \cite{turn0academia25} further extend pretraining objectives to a unified traffic understanding and generation, showing big potential across diverse downstream applications.
Beyond sequence-based modeling, several works explore alternative structural representations to improve robustness under distribution shifts. Transformer variants with multi-instance or hierarchical designs \cite{MIETT2024} and graph-based flow modeling approaches \cite{turn0academia21} aim to capture richer intra-flow interactions and structural dependencies. \textit{Despite these advances, most existing methods still rely on relatively uniform modeling pipelines across heterogeneous traffic components, without distinguishing the treatment of different ingredients in encrypted traffic. This hinders traffic representations.} %motivating the need for more adaptive and modular architectures.

\subsection{Mixture-of-Experts and Conditional Modeling}

Mixture-of-Experts (MoE) is a classical conditional computation framework for partitioning representation capacity among specialized submodels, where a gating network dynamically routes inputs to a subset of expert networks \cite{Jacobs1991,Jordan1994}. This paradigm enables the model to focus representational power on distinct subregions of the input space while minimizing redundant computation.
MoE architectures have been revitalized for large-scale neural networks. Shazeer et al. \cite{Shazeer2017} introduced the sparsely‑gated MoE layer, which scales model capacity by activating only a few experts per input, enabling models with increased orders of magnitude parameters yet without a proportional increase in computation cost. Extensions such as GShard \cite{GShard2020} further demonstrates that conditional routing and automatic partitioning of experts support effective scaling of very large Transformer models. The Switch Transformer \cite{SwitchTransformer2021} refines sparse routing mechanics to improve load balancing and training stability, achieving scalable performance with simplified routing strategies.
MoE has also been adapted beyond language modeling into multi‑task and multi‑modal learning. Multi‑Gate Mixture‑of‑Experts (MMoE) explicitly learns task‑specific expert routing to share and specialize features across tasks \cite{MMoE2018}. In computer vision, sparse MoE variants such as Vision MoE (V‑MoE) aggregate conditional expert selection into vision transformer architectures, achieving scalability and competitive accuracy with reduced computation compared to dense models \cite{VMoE2022}.

\textit{Despite these advances, most existing MoE designs assume homogeneous input or task‑level specialization, while encrypted traffic, by contrast, exhibits intrinsic structural heterogeneity, with headers and payloads carrying fundamentally distinct features.} This motivates our TrafficMoE, a dual‑branch MoE design, where MoE encoders and routing signals are aligned with heterogeneous traffic components, enabling adaptive and context‑aware traffic representation learning.

\begin{figure*}[!t]
    \centering
    \includegraphics[width=\textwidth]{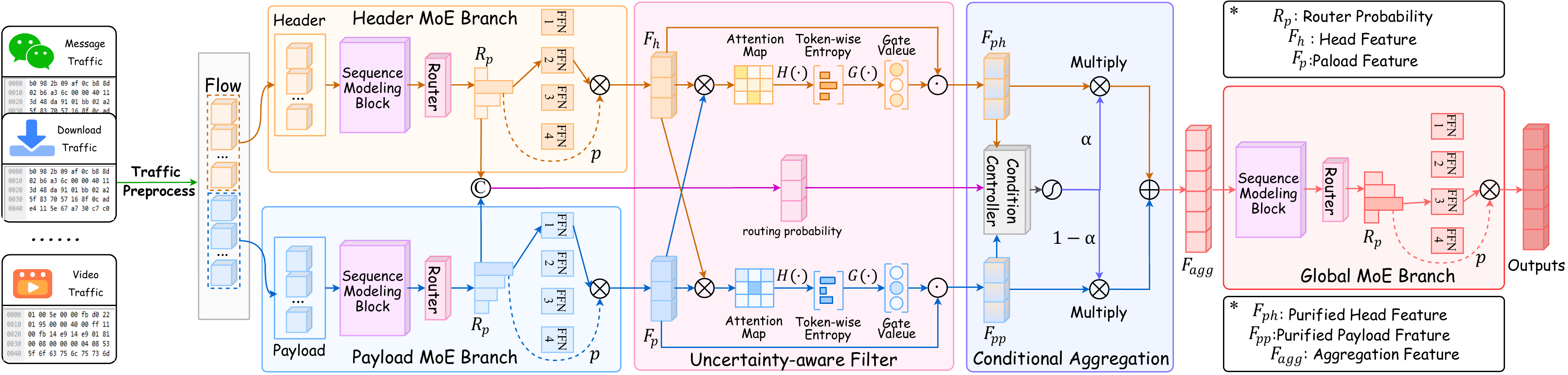}
    \caption{Overview of the TrafficMoE framework. The dual-branch MoE architecture explicitly models heterogeneous traffic components, i.e., the Header and Payload branches, which can capture intra-modality patterns, respectively. The uncertainty-aware filter suppresses unreliable tokens, i.e., noisy traffic components, and the conditional aggregation module fuses the final purified features across two modalities (header vs. payload) in a routing-guided manner for encrypted traffic classification.}
    \label{fig:overview}
    %\vspace{-1mm} % 可选，调整上下间距
\end{figure*}

\section{The Proposed TrafficMoE}

\subsection{Overview}

Encrypted network traffic exhibits intrinsic modality heterogeneity: the packet \textbf{headers} encode compact and structured protocol semantics with relatively stable patterns, whereas \textbf{payloads} encode long and high-entropy byte sequences with weak and noisy semantics. Directly applying a unified encoder to such inputs without distinction often leads to biased representations. To address this, we propose \textbf{TrafficMoE}, a heterogeneous-aware mixture-of-experts framework by following a \textit{Disentangle–Filter–Aggregate} paradigm, which is elaborated as Fig. \ref{fig:overview}. % designed under three principles: modality-decoupled sequence modeling, conditional expert specialization, and reliability-aware cross-modal integration. 
The framework comprises a Header branch and a Payload branch, each performing modality-specific sequence modeling followed by dedicated Mixture-of-Experts (MoE) layers, enabling conditional expert activation tailored to different properties of each modality. Representations from both branches are refined by an Uncertainty-aware Filtering (UF) module, in order to suppress unreliable tokens i.e., noisy traffic components based on attention-derived uncertainty, improving the stability of subsequent integration. The purified header and payload features in each branch are then aggregated via an implicit Conditional Aggregation (CA) mechanism that adaptively adjusts fusion behavior based on internal routing signals. To further enhance cross-modal adaptability and capture global traffic patterns, the aggregated representation is processed by a global MoE branch, producing a unified representation suitable for pre-training or task-specific fine-tuning.

\subsection{Traffic Preprocessing}\label{traffic_preprocessing}
\begin{figure}[!t]
    \centering
    \includegraphics[width=0.98\linewidth]{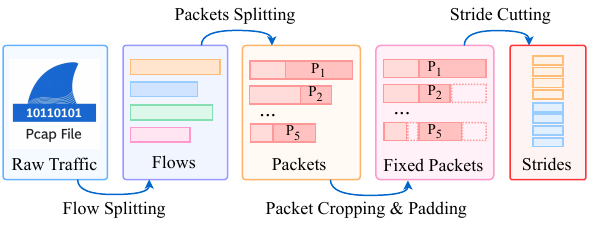}
  \caption{End-to-end preprocessing pipeline for encrypted traffic. Raw traffic is first segmented into flows using canonical 5-tuple session identification. These flows are further decomposed into packet-level units through packets splitting. Each packet undergoes byte-level cropping and zero-padding to produce fixed-dimensional header and payload segments. The resulting sequence of fixed packets is then partitioned into non-overlapping strides, enabling consistent, length-normalized inputs for subsequent neural processing while preserving temporal ordering and structural heterogeneity.}
    \label{fig:rawtraffic}
\end{figure}

The overall preprocessing pipeline is illustrated in Fig.~\ref{fig:rawtraffic}. 
Starting from the raw packet captures, traffic is progressively transformed into structured and length-normalized byte sequences suitable for heterogeneous modeling. 
The pipeline comprises four stages: flow-level aggregation, packet-level decomposition, byte-level cropping and padding, and stride-based segmentation. 
Each stage is designed to preserve temporal ordering and structural separability while producing stable tensor inputs for subsequent modality-specific processing.

\textbf{Stage 1: Flow Splitting.}
In network traffic analysis, a flow is defined as a sequence of packets sharing the same 5-tuple: source IP, destination IP, source port, destination port, and transport protocol. The raw traffic capture, denoted as $\mathcal{T}$, is first processed by aggregating packets into flows based on the canonical representation:
\begin{equation}
\mathcal{T} = \{F_1, F_2, \dots, F_K\}.
\label{eq:flowset}
\end{equation}
where each flow $F_k$ represents an ordered sequence of packets exchanged between a unique pair of endpoints. This splitting preserves session-level temporal dependencies essential for downstream sequence modeling.

\textbf{Stage 2: Packet Splitting.}
Each packet $P_i$ within a flow is decomposed into its constituent components:
\(
P_i = (H_i, B_i),
\)
where $H_i$ denotes the header bytes containing protocol metadata and control information, and $B_i$ represents the payload bytes carrying application data. Non-IP packets (e.g., ARP, DHCP) are removed to maintain semantic consistency, and link-layer headers are stripped to eliminate hardware-dependent artifacts and retain protocol-level information.

\textbf{Stage 3: Packet Cropping \& Padding.}
To produce fixed-size tensor inputs suitable for neural network processing, we constrain each flow to its first $M$ packets. Let $H_i = (h_{i,1}, h_{i,2}, \dots)$ and $B_i = (b_{i,1}, b_{i,2}, \dots)$ denote the raw byte sequences of the $i$-th header and payload, respectively. We construct fixed-length feature vectors $\tilde{H}_i \in \mathbb{R}^{N_h}$ and $\tilde{B}_i \in \mathbb{R}^{N_p}$ by retaining the initial bytes:
\begin{equation}
\tilde{H}_i = [h_{i,1}, \dots, h_{i, N_h}], \qquad \tilde{B}_i = [b_{i,1}, \dots, b_{i, N_p}].
\label{eq:crop}
\end{equation}
To ensure uniformity, zero-padding is applied where the sequence length falls short of the threshold, i.e., $h_{i,k}=0$ if $k > |H_i|$, and similarly for $B_i$.
Finally, the processed header--payload segments are concatenated in temporal order to form a structured byte sequence:
\begin{equation}
\mathbf{b} = [\tilde{H}_1, \tilde{B}_1, \dots, \tilde{H}_M, \tilde{B}_M],
\label{eq:concat}
\end{equation}
where its total length  $L_b = M(N_h+N_p)$.

\textbf{Stage 4: Stride Cutting.}
Instead of flattening $\mathbf{b}$ into a single long tensor that would impose artificial adjacency between unrelated byte regions, we segment the byte stream into fixed-length strides of size $L_s$:
\begin{equation}
\mathbf{s}_i = [b_{iL_s}, \dots, b_{(i+1)L_s - 1}], \quad 0 \le i < N_s,
\label{eq:stride}
\end{equation}
where $N_s = L_b / L_s$. This stride-based segmentation preserves temporal ordering and local contextual structures while normalizing the sequence length for efficient batch training. It further maintains structural separability, enabling header- and payload-specific processing pipelines to operate with their respective modality-aware inductive biases.

\subsection{Heterogeneous MoE Branches}\label{heter_moe}

Encrypted network traffic exhibits intrinsic modality heterogeneity: \textbf{headers} are short, structured, and semantically interpretable, while \textbf{payloads} are long, noisy, and highly variable. Optimizing a single shared encoder for both modalities often leads to biased representation, as the statistical and semantic properties of headers and payloads are fundamentally different. To explicitly account for this heterogeneity, TrafficMoE introduces two dedicated Mixture-of-Experts (MoE) branches—one for headers and another one for payloads. Each branch employs an identical architectural backbone but maintains independent expert sets and gating mechanisms, allowing experts to learn modality-specific distributions. This design leverages the core advantage of MoE: conditional expert activation enables flexible and context-aware modeling of distinct statistical regimes without requiring divergent architectures for each modality.

\textbf{General MoE Formulation.}
Each branch adopts a sparse Mixture-of-Experts formulation consisting of a content-aware gating module and an expert pool. 
Given an input sequence $X \in \mathbb{R}^{L \times D}$, the gating network first produces routing logits:
\begin{equation}
G = \mathrm{Gate}(X) = XW_g + b_g \in \mathbb{R}^{L \times E},
\end{equation}
where $W_g \in \mathbb{R}^{D \times E}$ and $b_g \in \mathbb{R}^{L\times E}$ are learnable parameters. 
Each row $G_\ell$ represents routing logits that measure the affinity between token $x_\ell$ and each expert.

For each token, only the Top-$K$ experts with the highest routing scores are selected. 
Formally, let $\mathcal{T}(x_\ell)$ denote the index set of the Top-$K$ experts for token $x_\ell$:
\begin{equation}
\mathcal{T}(x_\ell) = \mathrm{TopK}(G_\ell, K),
\end{equation}
where $\mathrm{TopK(\cdot)}$ means a top K selection function.
The routing weights are then normalized over the selected experts as:
\begin{equation}
R_{\ell,i} =
\begin{cases}
\frac{\exp(G_{\ell,i})}
{\sum_{j \in \mathcal{T}(x_\ell)} \exp(G_{\ell,j})},
& i \in \mathcal{T}(x_\ell), \\
0, & \text{otherwise}.
\end{cases}
\end{equation}

Each expert produces a transformation output:
\begin{equation}
F_i = \mathrm{Expert}_i(X), \quad i = 1, \dots, E,
\end{equation}
Then, the final MoE output is computed as a sparsely weighted aggregation:
\begin{equation}
F_\ell = \sum_{i \in \mathcal{T}(x_\ell)} 
R_{\ell,i} \cdot F_{\ell,i}.
\end{equation}

\textbf{Header Branch with Sequence Modeling and Top-$K$ Sparse MoE.}
Following the above MoE formulation, given structured header tokens $X_h \in \mathbb{R}^{L_h \times D}$ that encode stable and protocol-aligned semantics, 
we first apply a modality-specific sequence modeling block to capture contextual dependencies:
\begin{equation}
Z_h = \mathrm{SeqBlock}_h(X_h),
\end{equation}
where $\mathrm{SeqBlock}_h(\cdot)$ can be instantiated as multi-head self-attention or a state-space model (e.g., Mamba). 
This block enriches each header token with long-range contextual information while preserving structured protocol semantics.

The contextualized representations $Z_h$ are then routed through a header-specific sparse MoE module. 
The gating network produces routing logits as:
\begin{equation}
G_h = \mathrm{Gate}_h(Z_h) 
\in \mathbb{R}^{L_h \times E}.
\end{equation}

For each token $z_{h,\ell}$, only the Top-$K$ experts with the highest routing scores are selected:
\begin{equation}
\mathcal{T}_h(z_{h,\ell}) = 
\mathrm{TopK}(G_{h,\ell}, K),
\end{equation}
Then, the routing weights are normalized over the selected experts, formulated as:
\begin{equation}
R_{h,\ell,i} =
\begin{cases}
\frac{\exp(G_{h,\ell,i})}
{\sum_{j \in \mathcal{T}_h(z_{h,\ell})}
\exp(G_{h,\ell,j})},
& i \in \mathcal{T}_h(z_{h,\ell}) \\
0, & \text{otherwise}.
\end{cases}
\end{equation}

Each header expert produces a transformation output:
\begin{equation}
F^{(h)}_i = 
\mathrm{Expert}^{(h)}_i(Z_h),
\quad i = 1,\dots,E,
\end{equation}
Then, the final header representation is obtained via a sparse aggregation, shown as:
\begin{equation}
F_{h,\ell} =
\sum_{i \in \mathcal{T}_h(z_{h,\ell})}
R_{h,\ell,i} 
\cdot 
F^{(h)}_{\ell,i}.
\end{equation}

Essentially, each header expert models distinct protocol-related patterns, such as field interactions, control flags, protocol identifiers, etc. 
By performing Top-$K$ sparse routing over contextualized features, 
the header branch encourages structured and semantics-aware expert specialization 
while maintaining computational efficiency.

\textbf{Payload Branch with Sequence Modeling and Top-$K$ Sparse MoE.}
Payload tokens 
$X_p \in \mathbb{R}^{L_p \times D}$ 
exhibit high entropy, weak semantic locality, and substantial distributional variability due to encryption. 
Similarly, we first employ a payload-specific sequence modeling block:
\begin{equation}
Z_p = \mathrm{SeqBlock}_p(X_p),
\end{equation}
where $Z_p$ captures long-range statistical dependencies and implicit byte-level correlations within encrypted sequences.
The contextualized representations $Z_p$ are then routed through an independent sparse MoE module. 
The gating network produces routing logits:
\begin{equation}
G_p = \mathrm{Gate}_p(Z_p) 
\in \mathbb{R}^{L_p \times E}.
\end{equation}

For each payload token $z_{p,\ell}$, 
the Top-$K$ experts are selected:
\begin{equation}
\mathcal{T}_p(z_{p,\ell}) =
\mathrm{TopK}(G_{p,\ell}, K).
\end{equation}

Routing weights are normalized over the selected experts:
\begin{equation}
R_{p,\ell,i} =
\begin{cases}
\frac{\exp(G_{p,\ell,i})}
{\sum_{j \in \mathcal{T}_p(z_{p,\ell})}
\exp(G_{p,\ell,j})},
& i \in \mathcal{T}_p(z_{p,\ell}) \\
0, & \text{otherwise}.
\end{cases}
\end{equation}

Each payload expert produces a transformation output:
\begin{equation}
F^{(p)}_i =
\mathrm{Expert}^{(p)}_i(Z_p),
\quad i = 1,\dots,E.
\end{equation}

The final payload representation is computed as:
\begin{equation}
F_{p,\ell} =
\sum_{i \in \mathcal{T}_p(z_{p,\ell})}
R_{p,\ell,i}
\cdot
F^{(p)}_{\ell,i}.
\end{equation}

 In contrast to the header experts, payload experts tend to capture encrypted textures, statistical regularities, and noise-resilient patterns. The routing behavior in $R_p$ thus reflects distributional similarity rather than explicit semantics.
 Overall, the sequence modeling blocks and modality-specific MoE modules jointly disentangle \emph{contextual dependency modeling} from \emph{expert specialization}. This design enables TrafficMoE to independently encode structured protocol semantics and stochastic encrypted patterns, providing well-conditioned representations $F_h$ and $F_p$ for subsequent computations.

\subsection{Uncertainty-aware Filtering}

\begin{figure}[t]
    \centering
    \subfloat[Header Filtering ]{
        \includegraphics[width=0.47\linewidth]{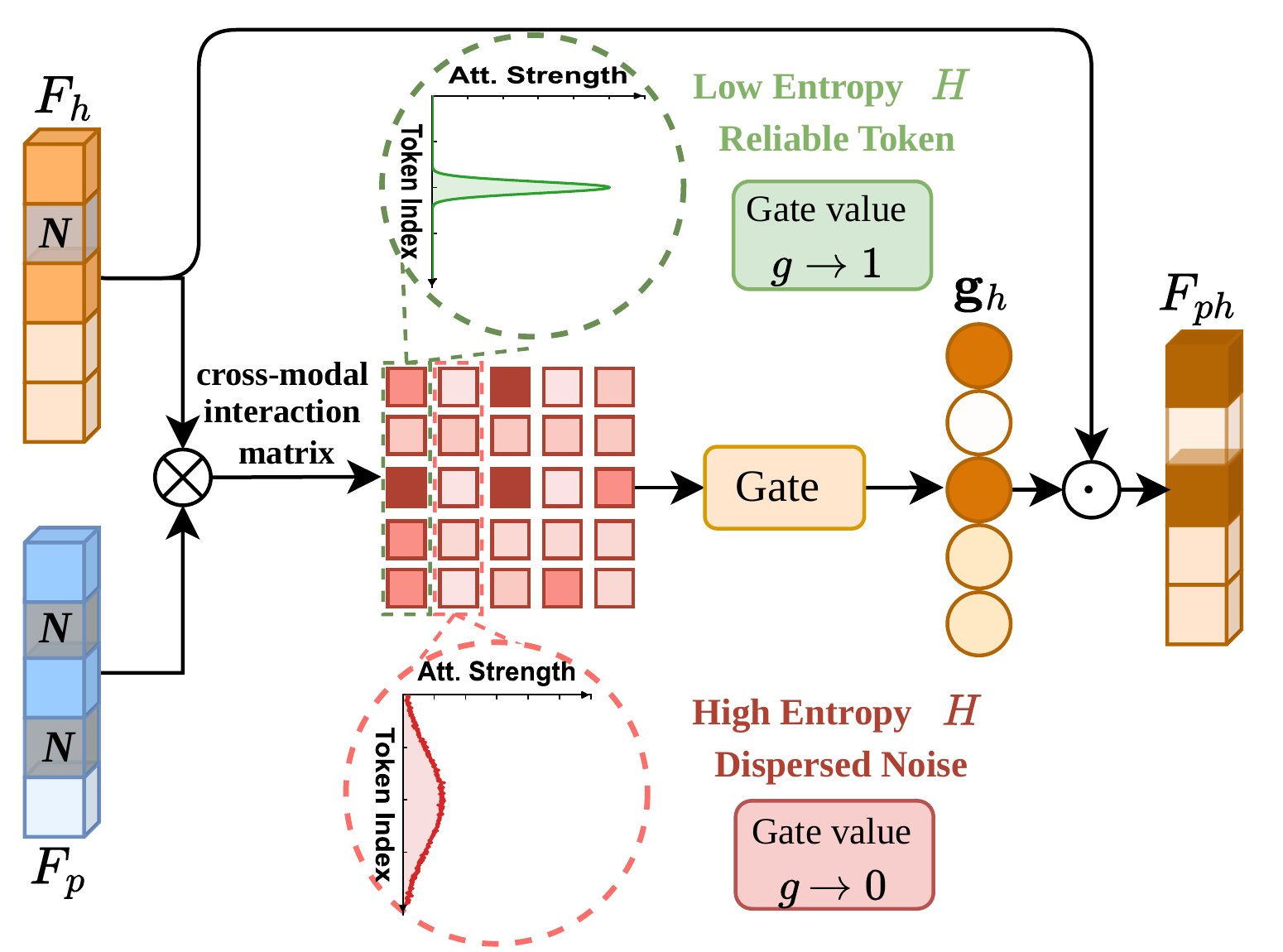}
        \label{subfig:header_filter}
    }
    \hfill
    \subfloat[Payload Filtering ]{
        \includegraphics[width=0.47\linewidth]{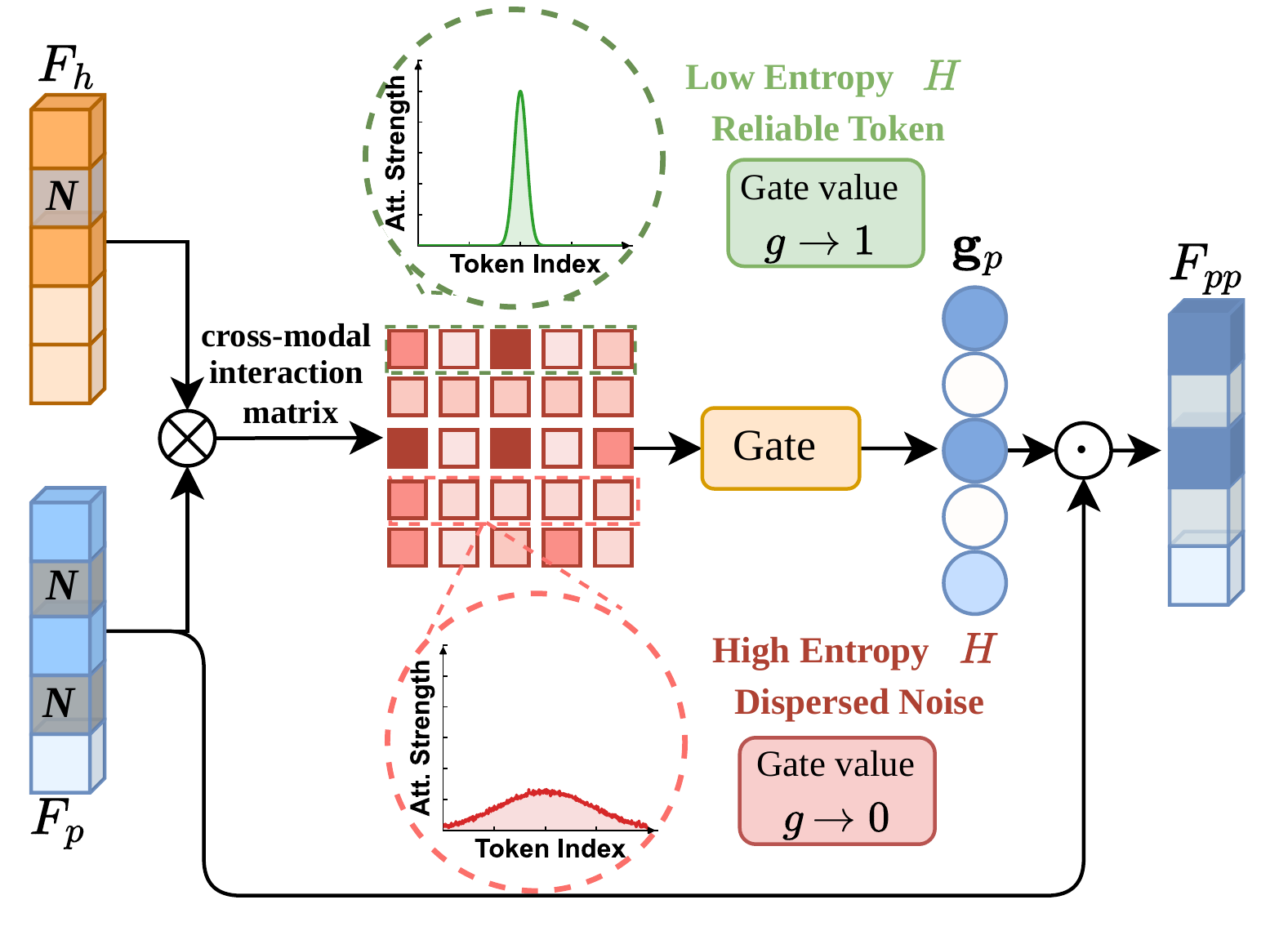}
        \label{subfig:payload_filter}
    }
    \caption{
    The basic idea of UF, which quantifies the alignment uncertainty via Shannon entropy ($H$). 
    Sharp distributions in green mean low entropy $H$ and identify reliable metadata to be retained (i.e., the filter $g \to 1$), while dispersed distributions in red mean high entropy $H$ and identify noisy components to be suppressed (i.e., the filter $g \to 0$).
    }
    \label{fig:UF_mechanism}
\end{figure}

Although the heterogeneous MoE branches generate modality-specific representations, encrypted traffic still contains unreliable, indiscriminative and noisy components, particularly within long and noisy payload sequences. Notably, MoE routing primarily encourages expert specialization rather than explicitly assess token reliability. As a result, tokens that exhibit unstable cross-modal interactions may continue to propagate noise to subsequent fusion layers. To address this issue, we introduce an \textbf{Uncertainty-aware Filter (UF)} that estimates token-level uncertainty based on cross-modal interaction between header and payload representations. Instead of relying on internal self-attention responses, UF enables a feature-level similarity computation between two modalities (i.e., headers and payloads), thereby capturing how each token consistently aligns with its cross-modal counterpart. Fig. \ref{fig:UF_mechanism} describes the basic idea of UF for header feature purification (Fig. \ref{fig:UF_mechanism}a) and payload feature purification (Fig. \ref{fig:UF_mechanism}b). We see that the UF module quantifies this alignment through the cross-modal interaction matrix's entropy, where \textit{low-entropy distributions identify reliable tokens to be retained, while high-entropy ones identify noisy components to be suppressed}.

\textbf{Token-wise Uncertainty Estimation.}
Let $F_{h,\ell} \in \mathbb{R}^{L_h \times d}$ and $F_{p,\ell} \in \mathbb{R}^{L_p \times d}$ denote the header and payload feature matrices, produced by their respective MoE branches. We compute a cross-modal interaction matrix:
\begin{equation}
A = \text{Softmax}\left( \frac{F_{h,\ell} F_{p,\ell}^\top}{\sqrt{d}} \right),
\label{attetion}
\end{equation}
where $A \in \mathbb{R}^{L_h \times L_p}$ measures the alignment strength between header tokens and payload tokens. Each row $A_i$ reflects how header token $h_i$ distributes its interaction over payload tokens.

We hypothesize that reliable tokens exhibit focused interaction patterns, while noisy or weakly informative tokens produce dispersed distributions. Accordingly, we quantify token-level uncertainty via entropy:
\begin{equation}
\begin{aligned}
H_h(i) &= - \sum_{j=1}^{L_p} A_{ij} \log(A_{ij} + \epsilon), \\
H_p(j) &= - \sum_{i=1}^{L_h} A_{ij} \log(A_{ij} + \epsilon),
\end{aligned}
\label{entropy_uncertainty}
\end{equation}
where $\epsilon$ ensures numerical stability and $i \in \{1,\ldots,L_h\}$ means the index of header tokens and $j \in \{1,\ldots,L_p\}$ denotes the index of payload tokens. Clearly, higher entropy indicates ambiguous or inconsistent cross-modal alignment, suggesting lower reliability. In contrast, lower entropy corresponds to a sharp interaction profile, implying stronger structural relevance across modalities.
The resulting uncertainty scores are used to softly suppress unreliable tokens through a learnable filtering mechanism before cross-modal fusion.

\textbf{Filter Weights Computation.}
To convert the above estimated uncertainty $H_h$ and $H_p$ in Eq.~(\ref{entropy_uncertainty}) into suppression weights, we introduce a learnable activation function:
\begin{equation}
\begin{aligned}
g_h^i &= \sigma(w H_h(i) + b), \quad i \in \{1,\ldots,L_h\}, \\
g_p^j &= \sigma(w H_p(j) + b), \quad j \in \{1,\ldots,L_p\},
\end{aligned}
\label{uncertainty_weight}
\end{equation}
 where $\sigma(\cdot)$ is the \textit{sigmoid} function, $w$ and $b$ are learnable weights.
 For a sequence of tokens, the token-wise activation scores are stacked as weight vector: 
 \begin{equation}
\begin{aligned}
\mathbf{g}_h &= [g_h^1, g_h^2, \ldots, g_h^{L_h}] \in \mathbb{R}^{L_h}, \\
\mathbf{g}_p &= [g_p^1, g_p^2, \ldots, g_p^{L_p}] \in \mathbb{R}^{L_p},
\end{aligned}
\label{gating_vector}
\end{equation}
The above computations enable the model to learn a soft thresholding behavior: \textit{larger entropy values are mapped to smaller scores, while tokens with low entropy are retained with higher weights}. During training, the function learns to correlate high uncertainty with negative gradients from the classifier, gradually shaping $w$ to enforce stronger suppression on noisy patterns. Thus, the activation score acts as a continuous reliability indicator that dynamically adjusts token contribution on a per-sample and per-token basis, rather than relying on a fixed global filtering rule.

\textbf{Feature Purification.}
The gating weights in Eq. (\ref{gating_vector}) are then multiplied element-wise to the MoE outputs of each branch. Denote header features by $F_h$ and payload features by $F_p$, the purified feature representations are computed as:
\begin{equation}
\begin{aligned}
F_{ph} &= \mathbf{g}_h \odot F_h, \\
F_{pp} &= \mathbf{g}_p \odot F_p.
\end{aligned}
\label{feat_purif}
\end{equation}
where $\mathbf{g}_h \in \mathbb{R}^{L_h}$ and $\mathbf{g}_p \in \mathbb{R}^{L_p}$ represent the token-level gating vectors for header and payload, respectively. This multiplicative modulation selectively suppresses noisy or unreliable activations while preserving structurally discriminative information. Crucially, the gating is performed at the token level rather than the sequence level, allowing the model to handle fine-grained intra-flow variability common in encrypted traffic. The purified representations $F_{ph}$ and $F_{pp}$ are then fed into the subsequent modules for computation. %provide cleaner, reliability-weighted inputs to the subsequent Conditional Fusion module, enabling it to focus on trustworthy cross-modal alignment rather than compensating for noise propagation.

\subsection{Implicit Conditional Aggregation}
To enable context-aware multi-modal integration while avoiding the rigidity of static fusion schemes, we propose an implicit \textbf{Conditional Aggregation (CA)} mechanism that adaptively modulates cross-modal interactions based on sample-specific traffic characteristics. Instead of introducing an explicit gating network, CA leverages the \textit{expert assignment probabilities} produced by the MoE routers in the header and payload branches as implicit conditional signals. These probabilities encode internal estimates of feature sparsity, modality saliency and structural heterogeneity, and thus serve as lightweight yet informative context descriptors without additional supervision.

\textbf{Context Encoding via Router Probabilities.}
Let $\mathbf{r}_h \in \mathbb{R}^{E}$ and $\mathbf{r}_p \in \mathbb{R}^{E}$ denote the soft expert-selection distributions from the header and payload MoE routers, respectively. We first project them into a shared conditional space:
\begin{equation}
\mathbf{c}_h = W_h \mathbf{r}_h, \qquad
\mathbf{c}_p = W_p \mathbf{r}_p,
\label{chcp}
\end{equation}
where $W_h$ and $W_p$ are learnable linear transformations. The resulting vectors are concatenated and normalized to form a unified conditional descriptor:
\begin{equation}
\mathbf{c} = \mathrm{Norm}([\mathbf{c}_h; \mathbf{c}_p]), 
\label{condition}
\end{equation}
Physically, this descriptor jointly captures intra-modal complexity and inter-modal complementarity.

\textbf{Context-Modulated Feature Aggregation.}
Given the intermediate representations $F_{ph}$ and $F_{pp}$ from the header and payload branches, in order to ensure stable relative scaling between modalities while preserving their distinct representational subspaces, we propose the conditional aggregation and modulation followed by feature concatenation:
\begin{equation}
\mathbf{F_{agg}} =
\Big[
\alpha(\mathbf{c}) \odot \phi(F_{ph}) \; ; \;
(1-\alpha(\mathbf{c})) \odot \psi(F_{pp})
\Big].
\label{Fagg}
\end{equation}
where $\phi(\cdot)$ and $\psi(\cdot)$ denote modality-specific transformations, and $\alpha(\cdot)$ is tractable conditioned on the descriptor $\mathbf{c}$ with softmax activation, $\phi(\cdot)$ and $\psi(\cdot)$ denote modality-specific linear projections that transform the header and payload representations into a unified feature space computed by:
\begin{equation}
\begin{aligned}
\alpha(\mathbf{c}) &= \mathrm{Softmax}(\mathbf{W}_c \mathbf{c} + \mathbf{b}_c), \\
\phi(F_{ph}) &= F_{ph} W_h, \\
\psi(F_{pp}) &= F_{pp} W_p.
\end{aligned}
\end{equation}
where $\mathbf{W}_c$ and $\mathbf{b}_c$ are learnable parameters that generate modality-aware weighting coefficients, while $W_h$ and $W_p$ denote learnable projection matrices for the header and payload representations, respectively.

\textbf{Adaptive Behavior and Advantages.}
Unlike the intuitive weighted summation, this formulation explicitly retains modality-specific information, %by explicitly preserving both branches in the fused representation, 
while allowing their contributions to be adaptively modulated based on traffic-specific context. The proposed CA mechanism introduces an \emph{implicit condition}, whereby fusion behavior is dynamically adapted using internal routing signals without additional gating supervision. By exploiting router probabilities that encode uncertainty, modality reliability and expert specialization patterns, CA achieves minimal parameter overhead, semantic consistency between expert routing and fusion, and improved robustness to heterogeneous and previously unseen encrypted traffic. Empirically, this design yields more expressive and stable representations than static fusion or fixed-weight fusion, particularly under modality-imbalanced conditions.

\begin{figure*}[!t]
    \centering
    %\vspace{-2mm}
    \includegraphics[width=0.90\textwidth]{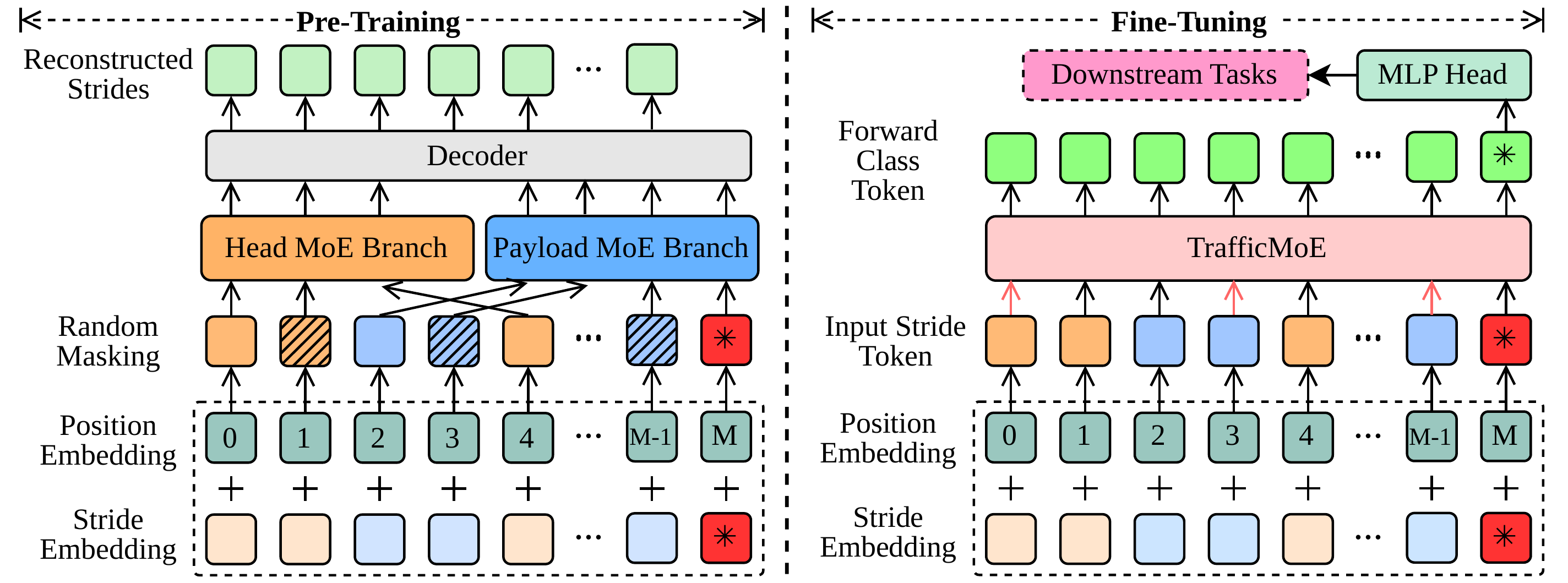}
    %\vspace{-2mm}
    \caption{Training pipeline of the proposed framework. In the pre-training stage, masked language modeling (MLM) is used to learn contextual representations from header and payload sequences without supervision. In the fine-tuning stage, the pretrained encoders are fine-tuned end-to-end on labeled traffic data through a standard cross-entropy classification objective.}
    \label{fig:pipeline}
    %\vspace{-3mm}
\end{figure*}

\subsection{Global MoE Branch}\label{global_moe}
The above CA module aggregates purified header and payload representations into a unified feature space. However, the fused representations may still exhibit substantial variability across traffic patterns. To further enhance cross-modal flexibility while maintaining computational efficiency and structural consistency with modality-specific branches, we introduce a \textbf{Global Sparse MoE Branch} for post-refinement.

Let $F_{agg} \in \mathbb{R}^{L_f \times D}$ denote the fused feature sequence. A sequence modeling block $\mathrm{SeqBlock}_g(\cdot)$ is first applied to capture contextual dependencies across the unified representation:
\begin{equation}
Z_f = \mathrm{SeqBlock}_g(F_{agg}),
\end{equation}
where $\mathrm{SeqBlock}_g(\cdot)$ can be instantiated as a multi-head self-attention module or a state-space model (e.g., Mamba), enabling modeling of long-range token interactions over fused cross-modal features.

Similarly, the contextualized representations $Z_f$ are then processed by a dedicated gating network:
\begin{equation}
G_g = \mathrm{Gate}_g(Z_f) \in \mathbb{R}^{L_f \times E}.
\end{equation}

For each token $z_{f,\ell}$, only the Top-$K$ experts with the highest routing scores are selected:
\begin{equation}
T_g(z_{f,\ell}) = \mathrm{TopK}(G_{g,\ell}, K).
\end{equation}

Then, the routing weights are normalized over the selected experts:
\begin{equation}
R_{g,\ell,i} =
\begin{cases}
\frac{\exp(G_{g,\ell,i})}
{\sum_{j \in T_g(z_{f,\ell})} \exp(G_{g,\ell,j})},
& i \in T_g(z_{f,\ell}), \\
0, & \text{otherwise}.
\end{cases}
\end{equation}

Each global expert generates a transformation output of the contextualized features:
\begin{equation}
F^{(g)}_i = \mathrm{Expert}^{(g)}_i(Z_f), \quad i = 1, \dots, E.
\end{equation}

The final representation is computed via sparse aggregation:
\begin{equation}
F_{g,\ell} =
\sum_{i \in T_g(z_{f,\ell})}
R_{g,\ell,i} \cdot F^{(g)}_{i,\ell}.
\end{equation}

Unlike the modality-specific branches that operate on structurally disentangled input, the global experts process already-fused cross-modal representations. By performing Top-$K$ sparse routing at this stage, the Global MoE Branch preserves conditional computation consistency across the entire architecture, while enabling high-level expert specialization over unified traffic semantics. This post-refinement further reinforces cross-modal features according to traffic-specific patterns and can be seamlessly exploited in both pre-training and fine-tuning stages.

\subsection{Training Pipeline of TrafficMoE}
The overall training pipeline of TrafficMoE is illustrated in Fig.~\ref{fig:pipeline}. 
The framework follows a two-stage optimization strategy consisting of self-supervised pretraining and supervised fine-tuning. 
In the pretraining stage, modality-specific encoders and MoE routing are initialized through masked language modeling (MLM) on unlabeled traffic flows. 
Subsequently, the entire architecture—including the MoE branches, UF module, and CA controller—is fine-tuned end-to-end on labeled encrypted traffic data.

\textbf{Pre-training Stage.}
To initialize modality-specific encoders and stabilize the heterogeneous MoE routing, we adopt a masked language modeling (MLM) objective tailored to encrypted traffic. Given a flow sample composed of a \textbf{header} token sequence $X_h=\{x_{h,i}|i=1,\cdots,L_h\}$ and a \textbf{payload} token sequence $X_p=\{x_{p,i}|i=1,\cdots,L_p\}$, we apply random masking to a portion of tokens within each sequence. Let $\tilde{X}_h$ and $\tilde{X}_p$ denote the masked sequences. %, and $\hat{X}_h$, $\hat{X}_p$ be the model's predictions.
Following the standard MLM formulation, the objective is to recover masked tokens based on their contextual dependencies. The loss of MLM for each branch is defined as:
\begin{equation}
\mathcal{L}_{MLM}^{(h)} = -\sum_{i \in \mathcal{M}_h} \log P(x_{h,i} \mid \tilde{X}_h),
\end{equation}
\begin{equation}
\mathcal{L}_{MLM}^{(p)} = -\sum_{i \in \mathcal{M}_p} \log P(x_{p,i} \mid \tilde{X}_p).
\end{equation}
where $\mathcal{M}_h$ and $\mathcal{M}_p$ represent the masked token indices for header and payload. The total pretraining loss is:
\begin{equation}
\mathcal{L}_{pre} = \mathcal{L}_{MLM}^{(h)} + \mathcal{L}_{MLM}^{(p)}.
\label{MLMloss}
\end{equation}

This pretraining strategy enables the header encoder to learn structural and protocol-specific patterns, while the payload encoder captures semantic and byte-level distributions. Moreover, the MoE architecture benefits from pretraining by allowing experts to specialize organically on different flow characteristics before downstream optimization. The UF module is jointly optimized during pretraining so that unreliable tokens receive appropriately low reliability coefficients, effectively regularizing early-stage representation learning.

\textbf{Fine-tuning Stage.}
For the downstream encrypted traffic classification task, both pretrained encoders and the CA module are optimized end-to-end. Given purified header features and purified payload features, the fusion controller generates sample-conditioned fusion coefficients that modulate the contributions of the two branches. The fused representation is fed into a classification head parameterized by $\theta_{cls}$.

Let $y$ denote the ground-truth traffic category, then the model outputs class probabilities:
\begin{equation}
P(y \mid X_h, X_p) = \text{softmax}\big( \theta_{cls}^\top F_{agg} \big).
\end{equation}

The fine-tuning objective is the standard cross-entropy loss:
\begin{equation}
\mathcal{L}_{cls} = - \log P(y \mid X_h, X_p).
\label{clsloss}
\end{equation}

During fine-tuning, all modules—including the MoE branches, UF reliability modulation, and the CA controller—are jointly optimized. This joint optimization allows the model to adapt the structural and semantic priors to the distributions of encrypted traffic categories in the downstream dataset, yielding robust and discriminative representations. 

To summarize, the training procedure of TrafficMoE is presented in \textbf{Algorithm 1}.

\section{Experiments}
\subsection{Experimental Setup}
\begin{algorithm}[t]
\caption{Training Procedure of TrafficMoE}
\label{alg:trafficmoe}
\begin{algorithmic}[1]

\REQUIRE Raw unlabeled traffic $\mathcal{T}_u$ and labeled traffic $\mathcal{T}_l$ with labels $Y$, the number $E$ of experts, and the parameter $K$ for Top-$K$ routing. 
\ENSURE Trained TrafficMoE model.

\STATE \textbf{Traffic preprocessing:}
Split raw traffic into flows and convert each flow into header tokens $X_h$ and payload tokens $X_p$ following the pipeline in Sec. \ref{traffic_preprocessing}.

\STATE \textbf{Stage I: Self-supervised Pre-training}

\FOR{each unlabeled training flow from $\mathcal{T}_u$}
    \STATE Apply random masking to obtain $\tilde{X}_h$ and $\tilde{X}_p$;
    \STATE \textcolor{green}{\textit{\#Disentangling}}
    \STATE Encode the header and payload sequences using modality-specific sequence blocks, respectively;
    \STATE Perform sparse MoE routing for header and payload branches as presented in Sec. \ref{heter_moe};
    \STATE \textcolor{green}{\textit{\#Filtering}}
    \STATE Estimate token uncertainty via cross-modal interaction using Eq. (\ref{attetion}) and Eq. (\ref{entropy_uncertainty});
    \STATE Compute filter weights using Eq. (\ref{uncertainty_weight}) and Eq. (\ref{gating_vector});
    \STATE Obtain purified features using Eq.(\ref{feat_purif});
    \STATE \textcolor{green}{\textit{\#Aggregation}}
    \STATE Perform conditional aggregation (CA) using Eq. (\ref{chcp}), Eq. (\ref{condition}) and Eq. (\ref{Fagg});
    \STATE Refine the fused representation using Global MoE branch as presented in Sec. \ref{global_moe};
    \STATE Update model parameters using MLM objective as formulated in Eq. (\ref{MLMloss}).
\ENDFOR

\STATE \textbf{Stage II: Supervised Fine-tuning}

\FOR{each labeled flow sample from $\mathcal{T}_l$}
    \STATE Forward propagate through TrafficMoE to obtain Aggregation feature $F_{agg}$;
    \STATE Optimize the classification objective in Eq. (\ref{clsloss}).
\ENDFOR

\STATE \textbf{return} Fine-tuned TrafficMoE model parameters.

\end{algorithmic}
\end{algorithm}

\textbf{Pre-training Datasets.}
In the self-supervised pre-training stage, we utilize approximately 30\,GB of unlabeled raw network traffic collected from three publicly available repositories: ISCX-VPN2016 (NonVPN portion)~\cite{ISCXVPN2016}, CICIDS2017 (Monday subset)~\cite{CIC-IDS2017}, and the WIDE backbone trace~\cite{WIDE}. These datasets jointly provide large-scale, diverse, and naturally distributed encrypted traffic, covering a wide range of real-world scenarios. To construct the training corpus, we extract the first 64 consecutive bytes from the Network layer of each packet, ensuring that the model captures essential structured information from the IP layer and those without relying on task-specific annotations.
The three sources contribute complementary traffic characteristics.

Specifically, ISCX-VPN2016 (NonVPN) offers application-level diversity under controlled conditions. CICIDS2017 (Monday) provides clean benign traffic generated in an enterprise-like environment without attacks. The WIDE backbone trace introduces large-volume, naturally occurred Internet backbone flows with substantial protocol and routing variability. The combination of structured laboratory traces and unstructured backbone traffic enhances the robustness and generalization of the pre-trained model across heterogeneous network environments.

\textbf{Fine-tuning Datasets.}
For supervised fine-tuning and evaluation, we employ six widely used encrypted traffic datasets: CSTNET-TLS~1.3~\cite{Et-bert}, ISCX-Tor2016~\cite{ISCXTor2016}, CIC-IoT2022~\cite{CICIoT2022}, USTC-TFC2016~\cite{USTC-TFC2016}, and ISCX-VPN2016~\cite{ISCXVPN2016} composed of two sub-datasets: ISCX-VPN(APP) and ISCX-VPN(Service). All flows are segmented into packets, and each packet is processed by extracting the first 64 bytes from the network-layer payload, ensuring consistent input format and preserving both protocol headers and encrypted payloads. Detailed statistical information for these datasets is summarized in Table~\ref{tab:dataset-summary}.
\begin{table}[htbp]
  \centering
  \begin{tabular}{lcc}
    \toprule
    Dataset & \# Sample & \# Category \\
    \midrule
    CSTNET-TLS 1.3    & 46,356 & 120 \\
    ISCX-Tor2016      & 14,569 & 16  \\
    CIC-IoT2022       & 22,634 & 6   \\
    USTC-TFC2016      & 50,677 & 20  \\
    ISCX-VPN (APP)    & 2,329  & 12  \\
    ISCX-VPN (Service)& 3,694  & 17  \\
    \bottomrule
  \end{tabular}
  \caption{Statistical Information of Fine-tuning Datasets}
  \label{tab:dataset-summary}
\end{table}

Specifically, CSTNET-TLS~1.3 contains large-scale TLS~1.3 encrypted traffic, representing contemporary application-layer encryption traffic. ISCX-Tor2016 consists of flows routed through the Tor anonymity network, exhibiting heavily obfuscated communication patterns. CIC-IoT2022 captures traffic from diverse IoT devices under both benign and malicious scenarios. USTC-TFC2016 includes balanced benign and malware flows with highly similar encrypted signatures, allowing for fine-grained classification evaluation. The ISCX-VPN (APP) subset aggregates traffic at the application level regardless of service type, while the ISCX-VPN (Service) subset focuses on service-level classification (e.g., Skype text, file transfer, and voice). These six datasets jointly provide a comprehensive benchmark spanning modern encrypted applications, anonymized routing, IoT ecosystems, malware flows, and VPN-obfuscated traffic.

During the fine-tuning stage, each traffic flow is represented by its first five packets, from which 64 bytes are extracted starting at the network layer for each packet. To protect sensitive information and reduce potential dataset-specific biases, the IP addresses and port numbers are randomized, and the TCP timestamp fields are normalized accordingly. Note that, all datasets are constructed from disjoint traffic sources to strictly prevent any form of data leakage.

\textbf{Evaluation Metrics.}
We evaluate the classification performance using four standard metrics, including \textit{Accuracy (AC)}, \textit{Precision (PR)}, \textit{Recall (RC)}, and the weighted \textit{F1-score (F1)}. Accuracy measures the overall proportion of correctly classified samples, while Precision and Recall quantify the reliability and completeness of class predictions, respectively. The weighted F1-score accounts for class imbalance by computing the harmonic mean of Precision and Recall, providing a comprehensive assessment across all traffic categories.

\begin{table*}[htbp]
    \small
    \centering
    \renewcommand{\arraystretch}{1.22}
    \tabcolsep=1.8mm
    \begin{tabular}{c|cccc|cccc|cccc}
    \toprule
    Dataset & \multicolumn{4}{c|}{ISCX-Tor2016} & \multicolumn{4}{c|}{CSTNET-TLS} & \multicolumn{4}{c}{CIC-IoT2022} \\
    \cline{1-13}
    Metric & AC & PR & RC & F1 & AC & PR & RC & F1 & AC & PR & RC & F1 \\
    \hline
    AppScanner\cite{APPscan}    & 0.9075 & 0.7728 & 0.8033 & 0.7848 & 0.7441 & 0.7232 & 0.6963 & 0.7023 & 0.8591 & 0.8858 & 0.7996 & 0.8288 \\
    BIND\cite{BIND}          & 0.9010 & 0.8582 & 0.8354 & 0.8439 & 0.4710 & 0.4315 & 0.4226 & 0.4189 & 0.7349 & 0.6754 & 0.6387 & 0.6435 \\
    CUMUL\cite{CUMUL}         & 0.7725 & 0.6463 & 0.6443 & 0.6401 & 0.5921 & 0.5528 & 0.5604 & 0.5493 & 0.7019 & 0.6746 & 0.7029 & 0.6687 \\
    DF\cite{DF}            & 0.7401 & 0.5918 & 0.5611 & 0.5492 & 0.5729 & 0.5398 & 0.5144 & 0.4933 & 0.2746 & 0.2140 & 0.1870 & 0.1647 \\
    FSNet\cite{FSNet2019}         & 0.6967 & 0.6159 & 0.6061 & 0.6028 & 0.7814 & 0.7670 & 0.7316 & 0.7311 & 0.8077 & 0.8250 & 0.8333 & 0.7804 \\
    GraphDApp\cite{GraphDapp}     & 0.7949 & 0.6391 & 0.6410 & 0.6383 & 0.7281 & 0.6964 & 0.6909 & 0.6890 & 0.7370 & 0.6721 & 0.7006 & 0.6767 \\
    Beauty\cite{Beauty}        & 0.3746 & 0.2691 & 0.2767 & 0.2251 & 0.2944 & 0.3219 & 0.2513 & 0.2324 & 0.1356 & 0.0349 & 0.0764 & 0.0296 \\
    ET-BERT\cite{Et-bert}       & 0.8123 & 0.7249 & 0.8276 & 0.7453 & 0.7993 & 0.7832 & 0.7689 & 0.7700 & 0.8603 & 0.8297 & 0.8255 & 0.8244 \\
    YaTC\cite{YATC}          & 0.9175 & 0.7725 & 0.7333 & 0.7405 & 0.8391 & 0.8364 & 0.8101 & 0.8140 & 0.8448 & 0.8656 & 0.8074 & 0.8048 \\
    TrafficFormer\cite{trafficformer} & 0.8669 & 0.7545 & 0.7460 & 0.7472 & 0.7982 & 0.7883 & 0.7736 & 0.7704 & 0.8725 & 0.8487 & 0.8343 & 0.8288 \\
    FlowletFormer\cite{flowletformer} & 0.9215 & 0.9263 & 0.9043 & 0.9116 &
                0.8605 & 0.8578 & 0.8445 & 0.8473 &
                0.9109 & 0.8905 & 0.8866 & 0.8859 \\
    \hline
    \textbf{TrafficMoE} & \textbf{0.9765} & \textbf{0.9768} & \textbf{0.9765} & \textbf{0.9765} &
                     \textbf{0.8688} & \textbf{0.8711} & \textbf{0.8688} & \textbf{0.8685} &
                     \textbf{0.9270} & \textbf{0.9265} & \textbf{0.9270} & \textbf{0.9265} \\
    \bottomrule
    \end{tabular}
    \vspace{4pt}
    \caption{Comparison results on ISCX-Tor2016, CSTNET-TLS, and CIC-IoT2022 datasets.}
    \label{tab:metrics-cross-dataset}
\end{table*}

\begin{table*}[htbp]
    \small
    \centering
    \renewcommand{\arraystretch}{1.22}
    \tabcolsep=1.8mm
    \begin{tabular}{c|cccc|cccc|cccc}
    \toprule
    Dataset & \multicolumn{4}{c|}{USTC-TFC} & \multicolumn{4}{c|}{ISCX-VPN (APP)} & \multicolumn{4}{c}{ISCX-VPN (Service)} \\
    \cline{1-13}
    Metric & AC & PR & RC & F1 & AC & PR & RC & F1 & AC & PR & RC & F1 \\
    \hline
    AppScanner\cite{APPscan}    & 0.8585 & 0.9108 & 0.9034 & 0.8976 & 0.7945 & 0.6950 & 0.6975 & 0.6874 & 0.8681 & 0.8710 & 0.8435 & 0.8546 \\
    BIND\cite{BIND}          & 0.7945 & 0.7811 & 0.7061 & 0.7115 & 0.6951 & 0.6266 & 0.5415 & 0.5609 & 0.8345 & 0.7769 & 0.7714 & 0.7699 \\
    CUMUL\cite{CUMUL}         & 0.7173 & 0.5063 & 0.5812 & 0.5183 & 0.5480 & 0.4839 & 0.4615 & 0.4554 & 0.7099 & 0.6959 & 0.6893 & 0.6884 \\
    DF\cite{DF}            & 0.6452 & 0.4019 & 0.3685 & 0.3059 & 0.4935 & 0.2449 & 0.2592 & 0.2289 & 0.5018 & 0.4664 & 0.4773 & 0.3934 \\
    FSNet\cite{FSNet2019}         & 0.7558 & 0.8167 & 0.8407 & 0.8042 & 0.6316 & 0.4899 & 0.4833 & 0.4677 & 0.9087 & 0.9051 & 0.9054 & 0.9051 \\
    GraphDApp\cite{GraphDapp}     & 0.8750 & 0.8446 & 0.8501 & 0.8249 & 0.5703 & 0.5108 & 0.4872 & 0.4853 & 0.7500 & 0.7311 & 0.7678 & 0.7429 \\
    Beauty\cite{Beauty}        & 0.6682 & 0.4448 & 0.4369 & 0.3796 & 0.6169 & 0.3333 & 0.3139 & 0.2964 & 0.6416 & 0.5769 & 0.5842 & 0.5387 \\
    ET-BERT\cite{Et-bert}       & 0.9663 & 0.9711 & 0.9663 & 0.9666 & 0.7964 & 0.7332 & 0.7013 & 0.7066 & 0.8467 & 0.8496 & 0.8651 & 0.8393 \\
    YaTC\cite{YATC}          & 0.9712 & 0.9732 & 0.9712 & 0.9707 & 0.8214 & 0.7443 & 0.7265 & 0.7254 & 0.9010 & 0.8877 & 0.8800 & 0.8821 \\
    TrafficFormer\cite{trafficformer} & 0.9750 & 0.9789 & 0.9750 & 0.9746 & 0.7751 & 0.7488 & 0.6846 & 0.6962 & 0.8533 & 0.8445 & 0.8348 & 0.8279 \\
    FlowletFormer\cite{flowletformer} & 0.9650 & 0.9689 & 0.9650 & 0.9648 & 0.8480 & 0.8153 & 0.7641 & 0.7712 & \textbf{0.9400} & \textbf{0.9471} & \textbf{0.9277}  & \textbf{0.9364}  \\
    \hline
    \textbf{TrafficMoE} & \textbf{0.9788} & \textbf{0.9788} & \textbf{0.9788} & \textbf{0.9788} &
                     \textbf{0.8872} & \textbf{0.8874} & \textbf{0.8872} & \textbf{0.8871} &
                     0.9255 & 0.9261 & 0.9255 & 0.9261 \\
    \bottomrule
    \end{tabular}
    \vspace{4pt}
    \caption{Comparison results on USTC-TFC, ISCX-VPN (APP), and ISCX-VPN (Service) datasets.}
    \label{tab:metrics_ustc_iscxvpn}
\end{table*}

\textbf{Comparison Methods.}
To comprehensively evaluate the effectiveness of our TrafficMoE framework, we compare against a broad range of representative approaches spanning three main methodological categories.

1) \textit{Machine learning-based methods.}
This category includes AppScanner\cite{APPscan}), BIND\cite{BIND}, and CUMUL\cite{CUMUL}. These approaches rely on handcrafted statistical or flow-level features, followed by traditional classifiers such as SVMs, random forests, or k-NN models. Thus, the performance highly depends on feature engineering and they typically struggle to generalize across heterogeneous or encrypted traffic patterns.

2) \textit{Deep learning-based methods.}
We further include several deep neural network models including DF \cite{DF}, FSNet \cite{FSNet2019}, GraphDapp \cite{GraphDapp}, and Beauty\cite{Beauty}, that directly operate on raw packet sequences or intermediate flow representations. These methods leverage CNNs, RNNs, or graph neural network architectures to automatically extract hierarchical traffic features. While more robust than conventional machine learning models, they remain limited by the scale and diversity of supervised training data and often lack the ability to generalize to unseen or novel traffic data or categories.

3) \textit{Pretraining-based methods.}
Finally, we compare against state-of-the-art pretraining approaches, including ET-BERT \cite{Et-bert}, YaTC \cite{YATC}), TrafficFormer\cite{trafficformer}, and FlowletFormer\cite{flowletformer}. These frameworks firstly perform pre-training to learn universal traffic semantics in a self-supervised or weakly-supervised manner on large-scale unlabeled corpora, and are subsequently fine-tuned on specific downstream traffic classification datasets. To ensure fair comparisons among different methods, all pretraining-based baselines share the same pretraining datasets and fine-tuning datasets.

\subsection{Comparison with State-of-the-Art Methods}
%We compare the proposed \textbf{TrafficMoE} framework with a wide range of representative state-of-the-art approaches on six public encrypted traffic classification benchmarks, including ISCX-Tor2016, CSTNET-TLS, CIC-IoT2022, USTC-TFC, ISCX-VPN(APP), and ISCX-VPN(Service). The compared methods span traditional machine-learning-based models, deep-learning-based architectures, and recent pretraining-based frameworks, 
The results with detailed comparisons to baselines and state-of-the-art methods are summarized in Tables~\ref{tab:metrics-cross-dataset} and~\ref{tab:metrics_ustc_iscxvpn}.

As shown in Table~\ref{tab:metrics-cross-dataset}, we have the following observations.

1) \textbf{ISCX-Tor2016.}
On this dataset, TrafficMoE achieves a substantial performance improvement over all competing methods, reaching 97.65\% for both accuracy and F1-score. In contrast, traditional machine learning-based approaches such as AppScanner, BIND, and CUMUL exhibit limited discriminative capability due to their reliance on handcrafted features. Deep learning-based models, including DF and FSNet, further struggle to capture the complex traffic patterns induced by Tor obfuscation. While pretraining-based methods such as FlowletFormer demonstrate strong performance, TrafficMoE still delivers a clear margin of improvement, highlighting the effectiveness of heterogeneity-aware expert routing architecture under highly anonymized traffic conditions.

2) \textbf{CSTNET-TLS.}
For this dataset that contains diverse TLS-encrypted application traffic, TrafficMoE consistently outperforms all baselines with an F1-score of 86.85\%. Compared with feature-based and supervised deep models that suffer from protocol and application variability, TrafficMoE exhibits superior robustness. Comparing against strong pretraining-based baselines such as ET-BERT, YaTC, and FlowletFormer, our method yields noticeable gains, indicating that explicitly modeling the heterogeneous traffic components for headers and payloads respectively via disentangling %and selectively filtering unreliable tokens 
is critical for TLS traffic understanding.

3) \textbf{CIC-IoT2022.}
On this large-scale and highly imbalanced IoT traffic dataset, TrafficMoE attains the best overall performance with an F1-score of 92.65\%. The classical deep models, such as DF, experience severe performance degradation due to their limited capacity to handle diverse IoT behaviors. In contrast, TrafficMoE effectively leverages expert specialization to disentangle heterogeneous IoT traffic patterns, resulting in significant improvements over both supervised and pretraining-based baselines.

As shown in Table~\ref{tab:metrics_ustc_iscxvpn}, we have the following observations.

1) \textbf{USTC-TFC.}
This dataset poses a challenging fine-grained traffic classification problem with subtle inter-class differences. TrafficMoE achieves an F1-score of 97.88\%, outperforming those strong pretraining-based models such as TrafficFormer and FlowletFormer. This  demonstrates the superiority of MoE in processing such data, and that uncertainty-aware filtering and conditional aggregation can effectively suppress noisy payload segments and enhance discriminability in fine-grained encrypted traffic scenarios.

2) \textbf{ISCX-VPN (APP).}
On this dataset that focuses on application-level VPN traffic classification, TrafficMoE achieves the highest F1-score of 88.71\%. While pretraining-based approaches already outperform traditional and supervised deep models, the performance is still constrained by modality noise in VPN tunneling. TrafficMoE mitigates this issue by dynamically allocating experts and adaptively weighing the contributions of different modalities (i.e., header vs. payload), bringing consistent gains for all evaluation metrics.

3) \textbf{ISCX-VPN (Service).}
For this dataset that emphasizes service-level semantics under VPN encryption, FlowletFormer achieves the state-of-the-art performance. Nevertheless, TrafficMoE remains highly competitive, achieving an F1-score of 92.61\%, which is comparable to FlowletFormer. This observation suggests that while service-level traffic may benefit from fine-grained temporal modeling, TrafficMoE still provides robust and stable performance by jointly exploiting heterogeneous representations and uncertainty-aware filtering.

\subsection{Ablation Studies}

%To comprehensively evaluate the contribution of each part in , 
We conduct a widespread ablation experiments for TrafficMoE by covering \textit{heterogeneous MoE structure}, \textit{key components}, and \textit{expert number}. %These analyses aim to rigorously validate the effectiveness of each component and provide deeper insights into the sources of performance improvements in the proposed framework.

\begin{table}[t]
\centering
\caption{Ablation on Heterogeneous MoE Structure.}
\label{tab:moe_ablation}
\resizebox{0.9\linewidth}{!}{
\begin{tabular}{l|c c}
\hline
\textbf{Method} & \textbf{Accuracy (\%)} & \textbf{F1 (\%)} \\
\hline
Heterogeneous MoE (Full) & 97.65 & 97.65 \\
Homogeneous MoE          & 92.21 & 92.32 \\
Header-only              & 75.65 & 75.65 \\
Payload-only             & 45.12 & 45.22 \\
\hline
\end{tabular}
}
\end{table}

\textbf{Ablation on Heterogeneous MoE Structure.}
We first investigate the impact of heterogeneous modeling in TrafficMoE by comparing with homogeneous and single-modality variants. The results are reported in Table~\ref{tab:moe_ablation}.

We observe that the full heterogeneous MoE achieves the highest performance, with 97.64\% accuracy and 97.44\% F1-score. When the heterogeneous design is replaced by a homogeneous MoE, where all experts share identical structures and process the mixed inputs of headers and payloads, the performance drops by more than 5\% on both metrics. This degradation indicates that simply increasing model capacity via multiple experts is insufficient; instead, explicitly assigning experts to structurally distinct traffic components w.r.t. headers and payloads is critical for effective representation learning.

More severe performance degradation is observed in the single-modality settings. The Header-only variant retains partial discriminative capability, reflecting that the protocol headers encode relatively stable structural cues even under encryption. In contrast, the Payload-only variant exhibits a dramatic performance collapse. This result suggests that the raw encrypted payload bytes are dominated by noisy and weakly discriminative patterns when modeled in isolation, making them unsuitable for standalone classification.

These above results demonstrate the performance gains of TrafficMoE stem from its heterogeneous expert design rather than the MoE architecture alone. By decoupling header-level structural information and payload-level encrypted content into dedicated experts, the model is able to better exploit complementary semantics while suppressing modality-specific noise. This heterogeneous formulation is therefore essential for robust encrypted traffic classification.

\begin{figure*}[t]
    \centering
    \subfloat[Header MoE]{
        \includegraphics[width=0.30\linewidth]{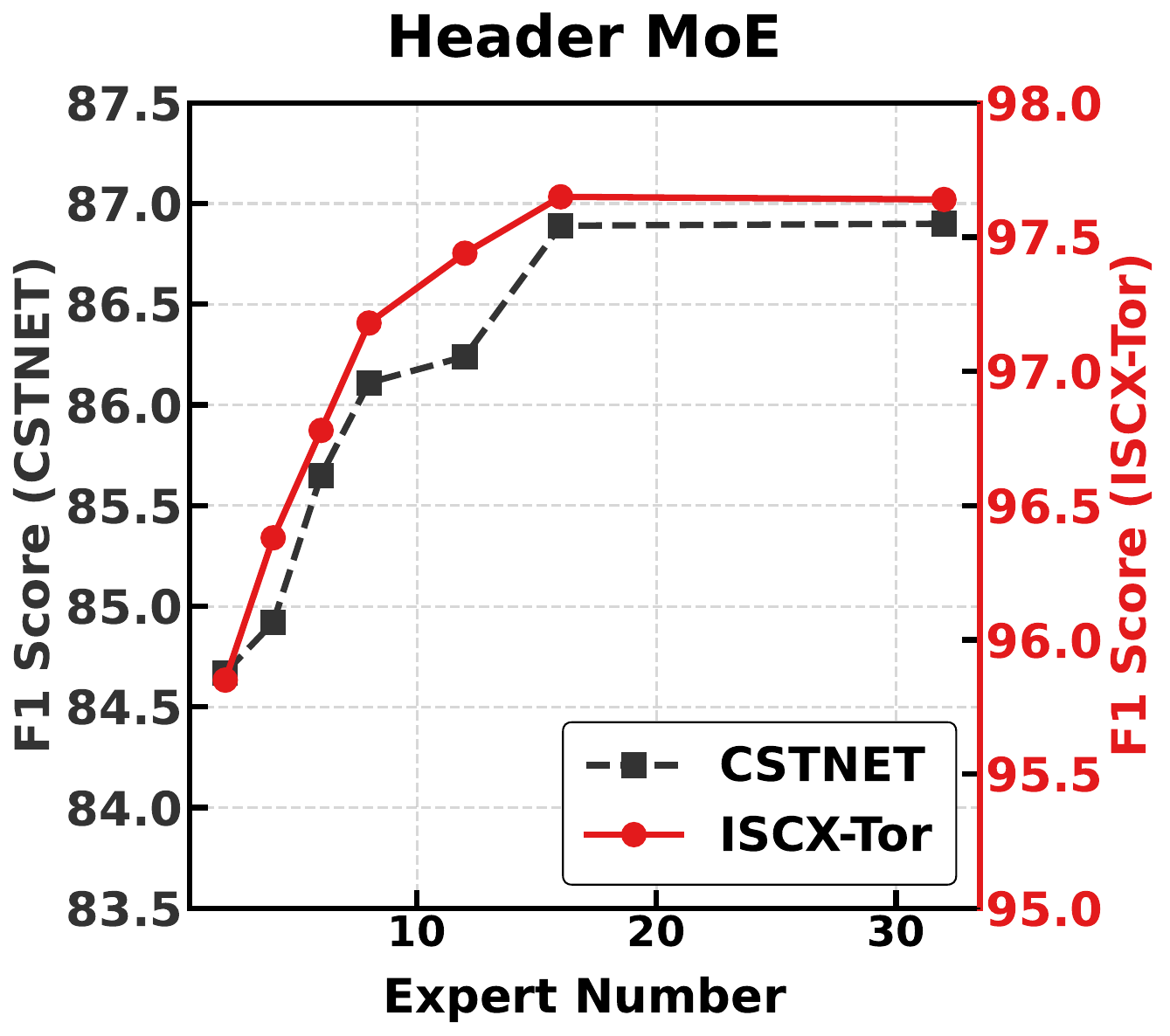}
        \label{subfig:ablation_tor}
    }
    \hfill
    \subfloat[Payload MoE]{
        \includegraphics[width=0.30\linewidth]{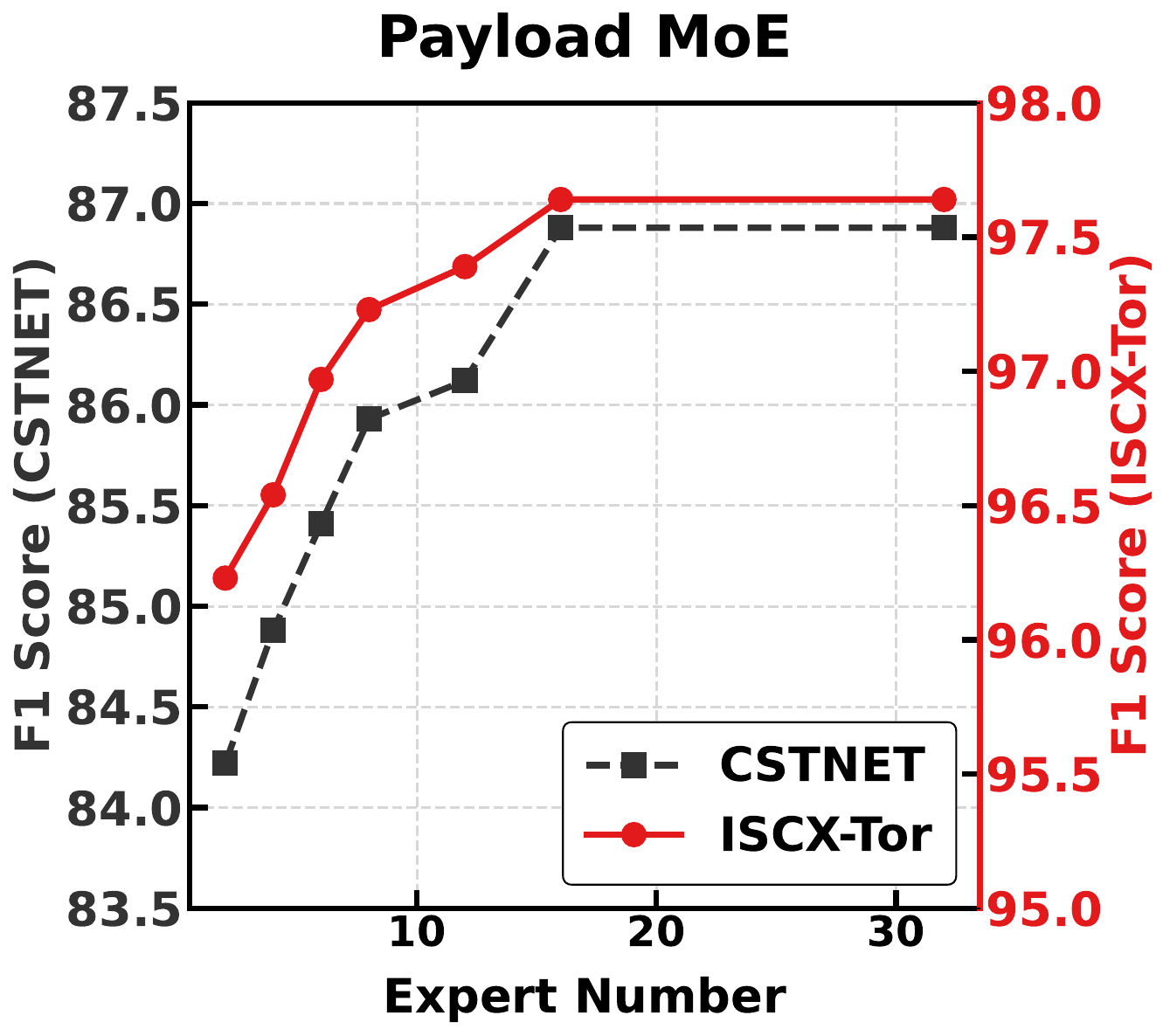}
        \label{subfig:ablation_cstnet}
    }
    \hfill
    \subfloat[Global MoE]{
        \includegraphics[width=0.30\linewidth]{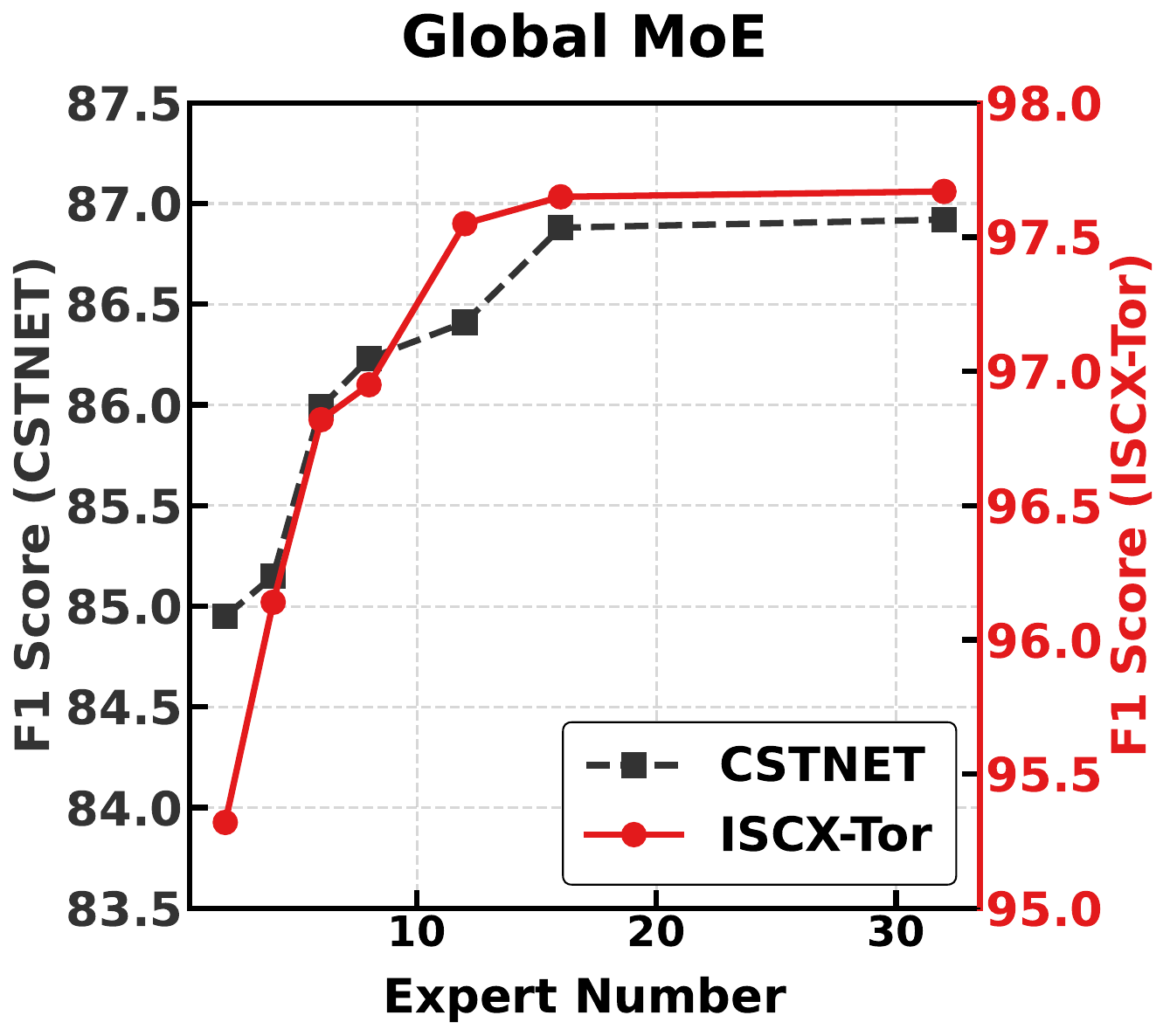}
        \label{subfig:ablation_cstnet}
    }
    \caption{Impact of the expert number in MoE for Header branch, Payload branch and the Global module, respectively, on ISCX-Tor2016 and CSTNET-TLS datasets. Clearly, MoE further improves representations.}
    \label{fig:expert_number_ablation}
\end{figure*}

\textbf{Ablation on Several Key Components.}
Table~\ref{tab:udf-ablation} further investigates the impact of key functional components in TrafficMoE, including \textit{uncertainty-aware filter (UF)}, \textit{conditional aggregation (CA)}, \textit{cross-modal interaction module} and \textit{pre-training}. Removing the UF module leads to a noticeable performance drop, indicating that entropy-based token filtering plays an essential role in suppressing redundant or unreliable tokens. Disabling the CA module also degrades performance, suggesting that adaptive modality weighting and feature aggregation is more effective than static fusion strategies.
Further, when the cross-modal interaction module is removed, the model still maintains relatively strong performance, but with a consistent degradation compared to the full model. This implies that explicit cross-modal alignment enhances fine-grained discrimination, especially under noisy or ambiguous traffic patterns. Finally, a drastic performance collapse is witnessed without pre-training, demonstrating that large-scale self-supervised pre-training is indispensable for learning transferable encrypted traffic representations.

Overall, these ablation results jointly validate that TrafficMoE benefits from a tightly coupled design combining heterogeneous expert modeling, uncertainty-aware filtering, conditional aggregation, and large-scale pre-training.

\begin{table}[t]
\centering
\caption{Ablation study of the  TrafficMoE. 
Accuracy and F1-score are reported under different filtering configurations.}
%\vspace{1mm}
\resizebox{0.90\linewidth}{!}{
\begin{tabular}{l|c c}
\hline
\textbf{Method} & \textbf{Accuracy (\%)} & \textbf{F1 (\%)} \\
\hline
TrafficMoE         & 97.65 & 97.65 \\
\hline
 w/o UF                       & 95.69 & 95.69 \\
 w/o CA                         & 96.33 & 96.33 \\
 w/o Cross-Modal Interaction         & 96.87 & 96.87 \\
 w/o Pre-training                        & 73.25 & 73.25 \\

\hline

\end{tabular}
}
\label{tab:udf-ablation}
\end{table}

\textbf{Impact of Expert Number.}
We further investigate the sensitivity of TrafficMoE to the number of experts in different branches, including the Header, Payload, and Global branches. Fig.~\ref{fig:expert_number_ablation} reports the variation of F1-score with respect to the expert number on ISCX-Tor2016 and CSTNET-TLS datasets.

On ISCX-Tor2016, all three branches as shown in Fig.~\ref{fig:expert_number_ablation} (a), (b) and (c) exhibit rapid performance improvements when the number of experts increases from 2 to 8, followed by a smoother convergence beyond 16 experts. Notably, the Payload branch shows the most pronounced performance gain in the low-expert regime, reflecting the high heterogeneity and noisy characteristics of Tor-encrypted payloads, which benefit substantially from increased expert diversity. In contrast, the Header branch converges more quickly, indicating that header-level patterns are relatively more structured. %require fewer experts to achieve stable performance.
A similar trend is observed on CSTNET-TLS. The F1-score steadily improves with increasing expert numbers across all branches. The Global branch demonstrates relatively stable performance, suggesting that global representations aggregate complementary information from both header and payload streams and are less sensitive to over-parameterization. Increasing the number of experts beyond 16 yields marginal improvements, highlighting a favorable trade-off between performance and efficiency.

Overall, increasing the number of experts consistently improves classification performance across all branches on both datasets, indicating that larger expert pools enable more fine-grained specialization and richer representation capacity. The performance gains gradually saturate as the number of experts increases, %, suggesting diminishing returns beyond a moderate model scale. 
suggesting that a moderate number of experts is sufficient to balance model capacity and efficiency, while excessive scaling may lead to redundant specialization without significant performance gains.

\subsection{Expert Behavior and Visualization Analysis}
\begin{figure}[t]
    \centering
    \subfloat[Early training stage]{
        \includegraphics[width=0.45\linewidth]{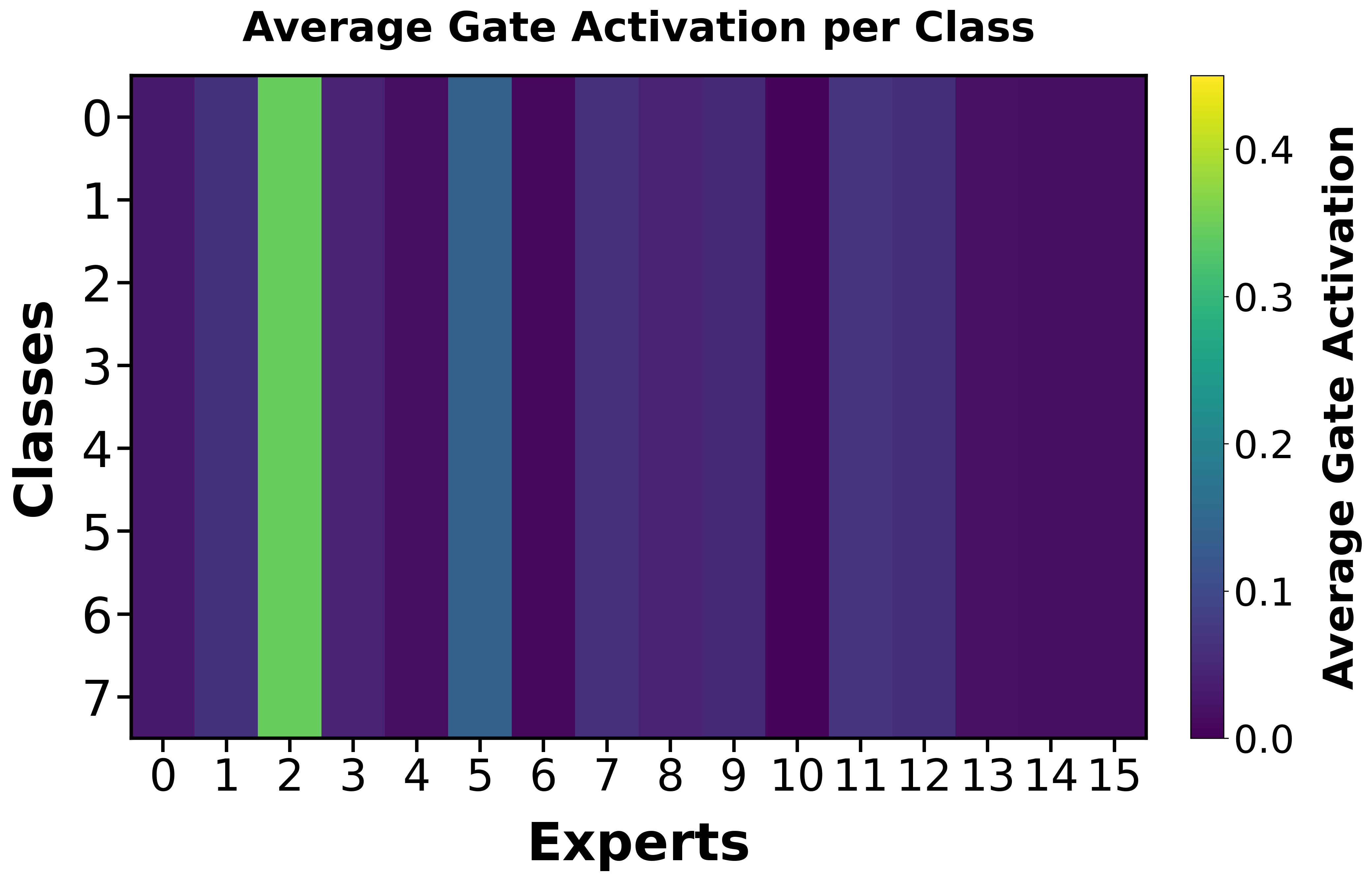}
        \label{subfig:expert_class_epoch0}
    }
    \hfill
    \subfloat[Converged stage]{
        \includegraphics[width=0.45\linewidth]{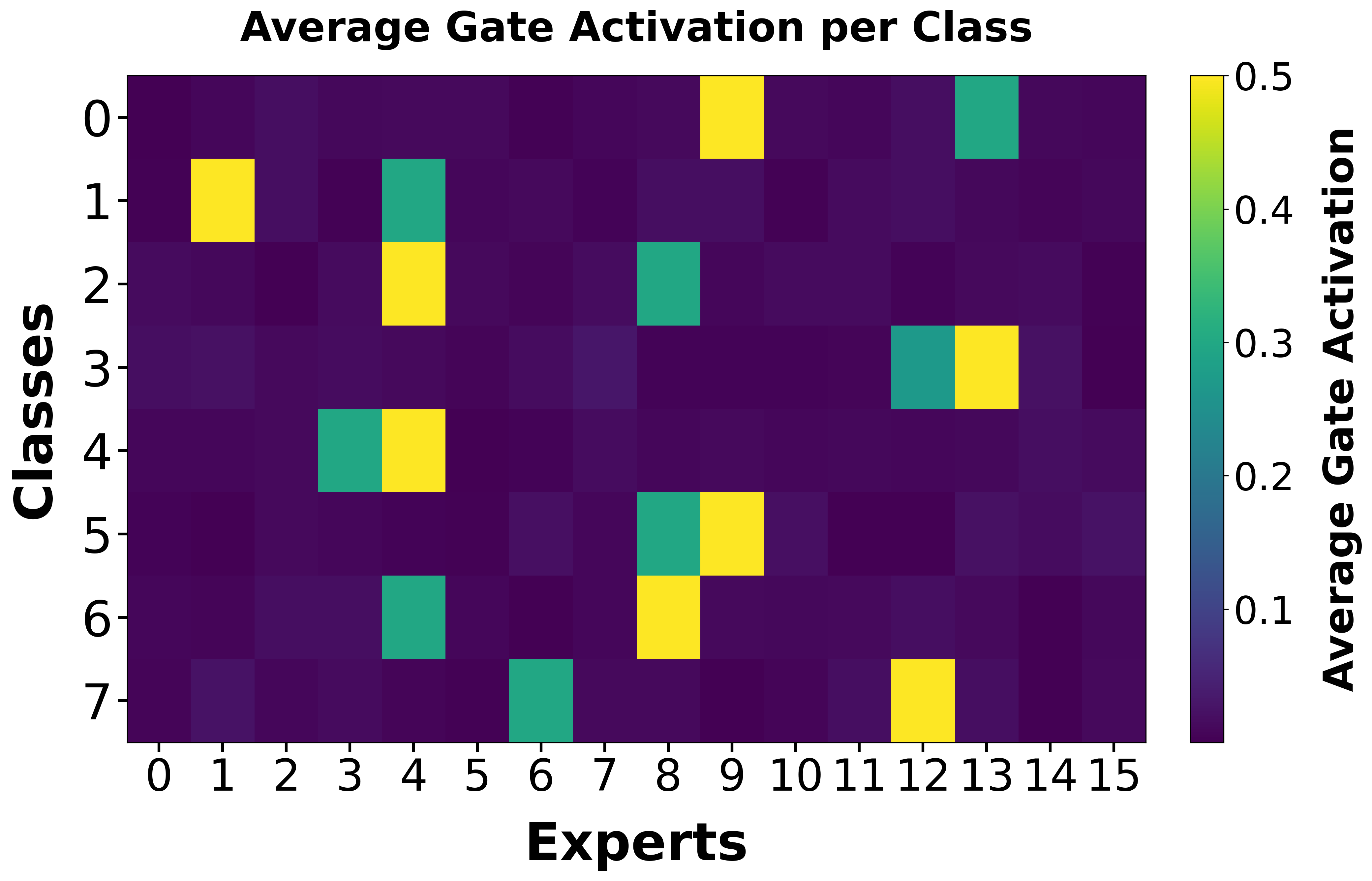}
        \label{subfig:expert_class_epoch120}
    }
    \caption{
    Class-wise expert behavior analysis of TrafficMoE.
    The average expert activation for each traffic class at the early training stage (Epoch 0) and converged stage (Epoch 120) is visualized, respectively.
    As training progresses, the routing distributions evolve from dispersed and class-agnostic patterns to structured and class-specific activations, indicating the expert specialization is aligned with semantic traffic categories.
    }
    \label{fig:expert_class_ablation}
\end{figure}

\textbf{Class-wise Expert Activation Behavior Analysis.}
To further investigate whether the proposed TrafficMoE learns meaningful expert specialization beyond performance gains, we analyze the class-wise expert activation patterns during training, i.e., \textit{expert behavior w.r.t. different classes}.
Fig.~\ref{fig:expert_class_ablation} visualizes the average routing weights of experts for each traffic class at the initial training stage (i.e., Epoch 0) and converged stage (i.e., Epoch 120).
\textit{At the early stage of training}, the routing distributions across experts are largely dispersed and exhibit weak class dependency.
Different traffic categories activate similar subsets of experts with comparable weights, indicating that the routing mechanism has not yet achieved structured specialization and behaves close to class-agnostic assignment.
This observation is consistent with randomly initialized gating functions, where expert selection is primarily driven by stochastic variations rather than semantic distinctions.
As training progresses, the activation patterns evolve into significantly more structured and class-aware distributions.
\textit{At the converged stage}, each traffic class consistently focuses its activation on a small subset of experts, while suppressing others, leading to clearly distinguishable ``expert–class" associations.
That is, different classes exhibit distinct expert preference profiles, demonstrating that the routing mechanism adaptively aligns experts with class-specific traffic characteristics.

These results provide strong evidence that TrafficMoE does not merely rely on increased model capacity, but instead learns semantically meaningful expert specialization through data-driven routing.
Such emergent class-aware expert activation supports the effectiveness of the heterogeneous MoE design in capturing diverse traffic patterns and contributes to the interpretability and robustness of the proposed framework.

\begin{figure}[t]
    \centering
    \subfloat[Before UF]{
        \includegraphics[width=0.45\linewidth]{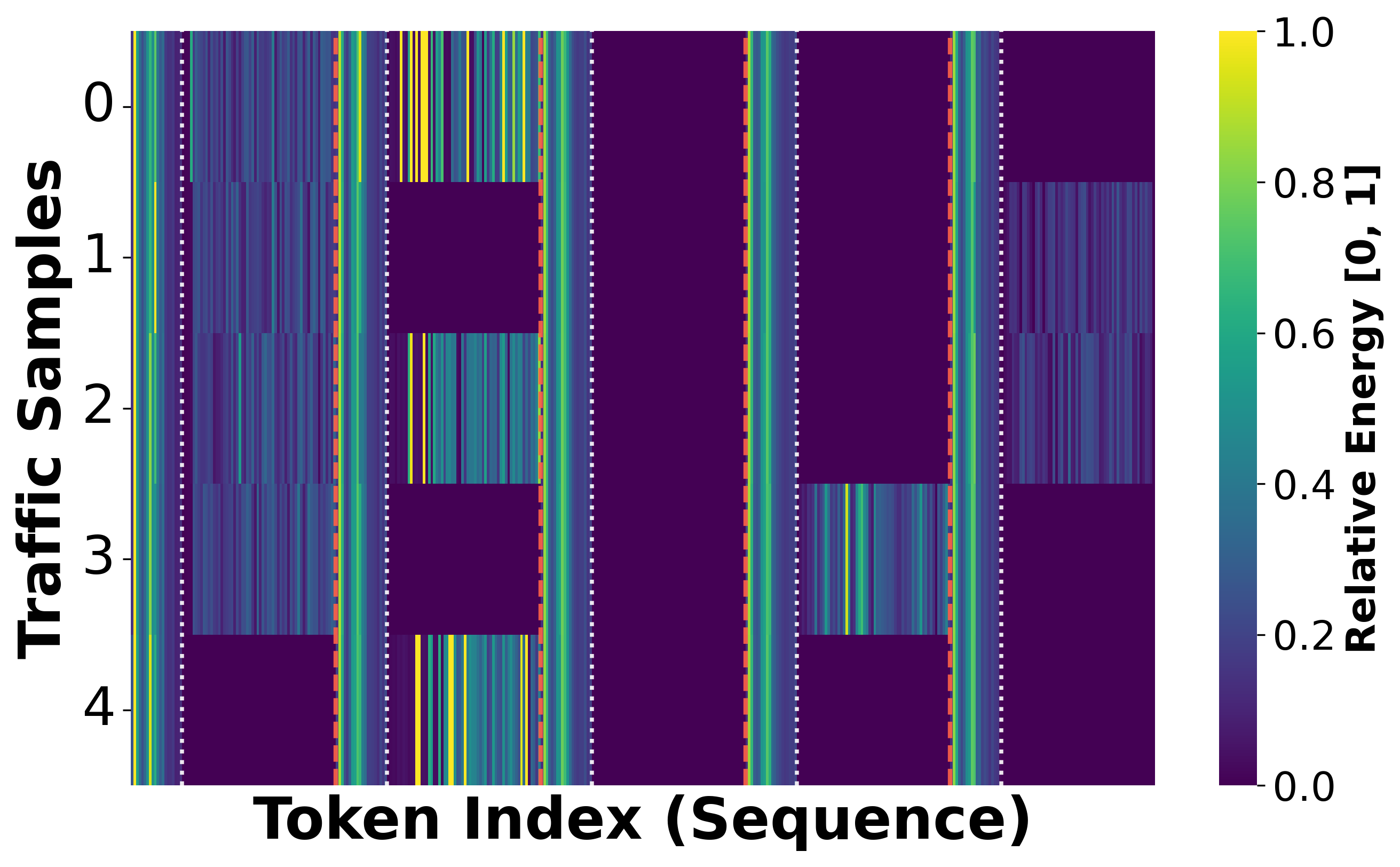}
        \label{subfig:feature_energy_before}
    }
    \hfill
    \subfloat[After UF]{
        \includegraphics[width=0.45\linewidth]{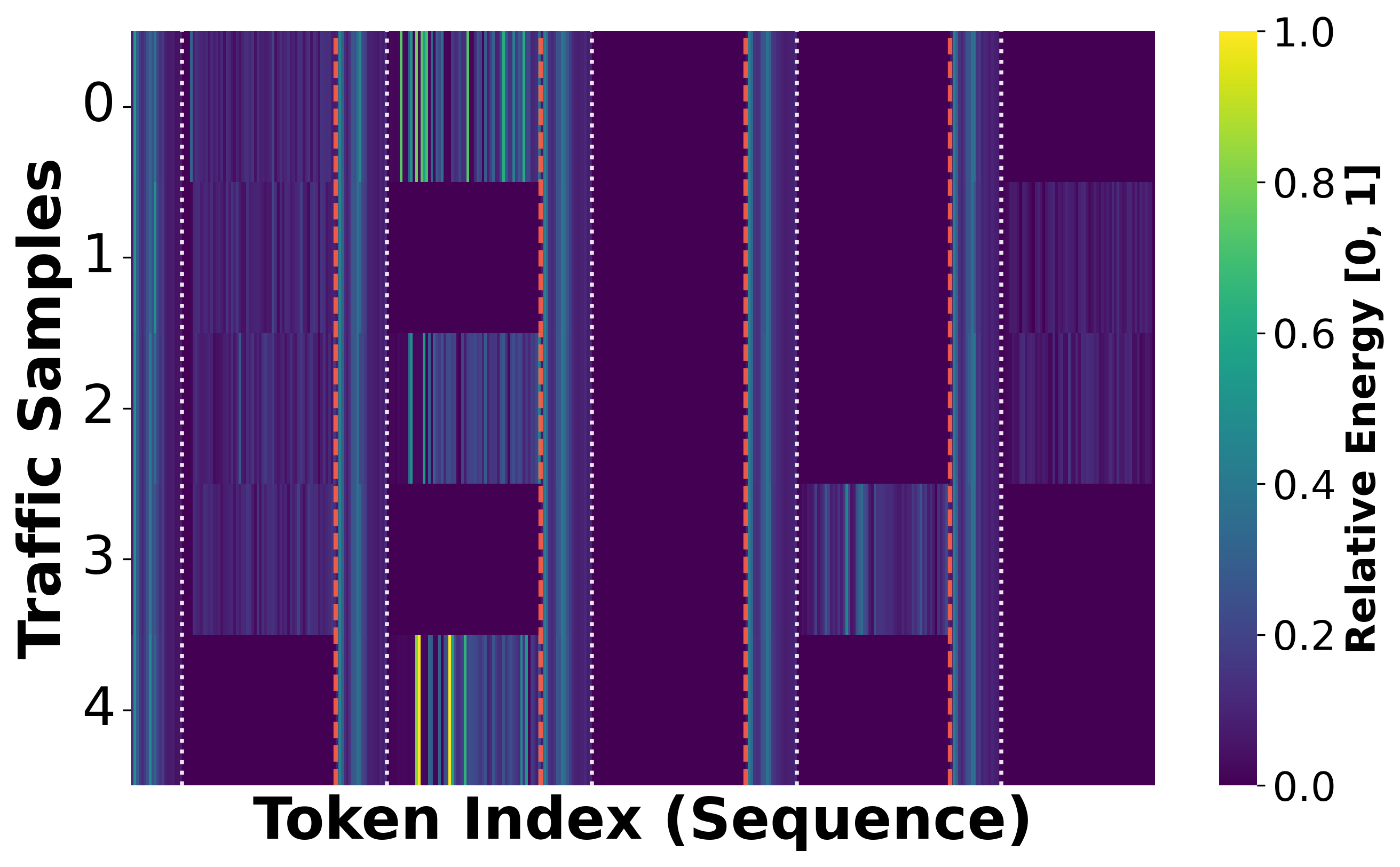}
        \label{subfig:feature_energy_after}
    }
    \caption{
    Visualization of UF module before and after filtering. 
    The heatmaps illustrate the relative feature magnitude of interleaved packet sequences across different traffic samples. The red dashed lines delineate the boundaries between individual network packets, while the white dotted lines separate the header and payload segments within each packet. After filtering, UF drastically suppresses the noisy components with high-energy activations in encrypted payload segments.
    %(a) Before filtering, the encrypted payload blocks exhibit high-energy activations that act as dispersed cross-modal noise. 
    %(b) After filtering, UF drastically suppresses these noisy components.%, while strictly preserving the reliable header metadata and sparse discriminative payload signatures. %This validates UF's capability to purify the feature space for subsequent expert routing.
    }
    \label{fig:UFvis}
\end{figure}

\textbf{Visualization of Uncertainty-aware Filtering.} 
To intuitively demonstrate the efficacy of the proposed UF module, we visualize the feature energy magnitude of token sequences before and after the filtering process via $L_2$ norm. As shown in Fig. \ref{fig:UFvis}, the input sequence preserves the physical temporal structure of raw traffic, interleaving header and payload patches across multiple packets. Before filtering, the encrypted payload segments exhibit dense, high-energy activations with noisy information. %These chaotic activations act as dispersed cross-modal noise, which can severely mislead the subsequent MoE expert routing.
After filtering, the high-entropy noisy components are drastically suppressed. Conversely, the reliable header metadata and highly discriminative signatures in payloads are strictly preserved. %This verifies that the UF module effectively purifies the feature.% by disentangling deterministic application patterns from pseudo-random encrypted noise.

\section{Conclusion and Future Work}

In this paper, we present a heterogeneous-aware mixture-of-experts (TrafficMoE) framework for encrypted traffic  interpretation, which follows a \textit{Disentangle-Filter-Aggregate} paradigm. Motivated by the intrinsic structural disparity between headers and payloads, we propose a dual-branch architecture with modality-specific MoE modules to enable conditional expert specialization without cross-modal interference. To further suppress unreliable tokens and noisy components, we propose an Uncertainty-aware Filtering (UF) mechanism that estimates token-level uncertainty from cross-modal interaction entropy, enabling both header and payload feature purification. An implicit Conditional Aggregation (CA) module, followed by a Global MoE branch, is then designed to adaptively aggregate purified representations and enhance cross-modal adaptability. 

%Extensive experiments on encrypted traffic benchmarks demonstrate that TrafficMoE achieves consistent improvements over state-of-the-art methods in both accuracy and robustness, while exhibiting clear expert specialization behavior across traffic categories. These results validate the effectiveness of heterogeneous-aware expert modeling and reliability-guided integration for encrypted traffic analysis.

Despite its effectiveness, several directions remain for future investigation. \textit{First}, current expert routing is learned in a data-driven manner without explicit structural priors. Incorporating protocol-level or temporal constraints into the routing process may further enhance interpretability and stability. \textit{Second}, extending the framework to continual or open-world traffic scenarios, where traffic distributions evolve over time, is challenging, unexplored and an promising direction. \textit{Finally}, exploring more principled uncertainty estimation strategies beyond entropy-based interaction measures may provide deeper theoretical grounding for reliability-aware modeling in encrypted environments.

\bibliographystyle{IEEEtran}
\bibliography{sample}

% \bibliographystyle{IEEEtran}

% \bibitem{ref1}
% {\it{Mathematics Into Type}}. American Mathematical Society. [Online]. Available: https://www.ams.org/arc/styleguide/mit-2.pdf

% \bibitem{ref2}
% T. W. Chaundy, P. R. Barrett and C. Batey, {\it{The Printing of Mathematics}}. London, U.K., Oxford Univ. Press, 1954.

% \bibitem{ref3}
% F. Mittelbach and M. Goossens, {\it{The \LaTeX Companion}}, 2nd ed. Boston, MA, USA: Pearson, 2004.

% \bibitem{ref4}
% G. Gr\"atzer, {\it{More Math Into LaTeX}}, New York, NY, USA: Springer, 2007.

% \bibitem{ref5}M. Letourneau and J. W. Sharp, {\it{AMS-StyleGuide-online.pdf,}} American Mathematical Society, Providence, RI, USA, [Online]. Available: http://www.ams.org/arc/styleguide/index.html

% \bibitem{ref6}
% H. Sira-Ramirez, ``On the sliding mode control of nonlinear systems,'' \textit{Syst. Control Lett.}, vol. 19, pp. 303--312, 1992.

% \bibitem{ref7}
% A. Levant, ``Exact differentiation of signals with unbounded higher derivatives,''  in \textit{Proc. 45th IEEE Conf. Decis.
% Control}, San Diego, CA, USA, 2006, pp. 5585--5590. DOI: 10.1109/CDC.2006.377165.

% \bibitem{ref8}
% M. Fliess, C. Join, and H. Sira-Ramirez, ``Non-linear estimation is easy,'' \textit{Int. J. Model., Ident. Control}, vol. 4, no. 1, pp. 12--27, 2008.

% \bibitem{ref9}
% R. Ortega, A. Astolfi, G. Bastin, and H. Rodriguez, ``Stabilization of food-chain systems using a port-controlled Hamiltonian description,'' in \textit{Proc. Amer. Control Conf.}, Chicago, IL, USA,
% 2000, pp. 2245--2249.

% \end{thebibliography}

\newpage

\vfill

\end{document}